\documentclass[prd,twocolumn,showpacs]{revtex4}
\usepackage{hyperref,amssymb,amsmath,mathrsfs,bm,graphicx}
\usepackage{graphicx,amsmath,amssymb,pgf}
\usepackage{enumerate}
\usepackage{color}
\begin{document}
\title{Extended two dimensional characteristic framework to study nonrotating black holes}
\author{{W. Barreto}}
\address{Centro de F\'\i sica Fundamental, Facultad de Ciencias, Universidad de Los
Andes, M\'erida 5101, Venezuela}
\date{\today}
\begin{abstract}
{We develop a numerical solver, that extends the computational framework considered in [Phys. Rev. D {\bf 65}, 084016 (2002)], to include scalar
perturbations of nonrotating black holes. The nonlinear Einstein-Klein-Gordon equations
for a massless scalar field minimally coupled to gravity are solved in two spatial dimensions (2D).
The numerical procedure is based on the ingoing
light cone formulation for an axially and reflection symmetric spacetime.
The solver is second order accurate and was validated in different ways.
We use for calibration an auxiliary 1D solver with the same initial and boundary conditions and the same evolution algorithm. We reproduce the quasinormal
modes for the massless scalar field harmonics $\ell = 0$, $1$ and $2$. For these same harmonics, in the linear approximation, we calculate the balance of energy between the black hole and the world tube. {As an example of nonlinear harmonic generation, we show the distortion of a marginally trapped two-surface approximated as a q-boundary and based upon the harmonic $\ell=2$.}  Additionally, we study the evolution of the $\ell = 8$ harmonic in order to test the solver in a spacetime with a complex angular structure. Further applications and extensions are briefly discussed.}
\end{abstract}
\maketitle
\section{Introduction}
{Relevant astrophysical applications of the characteristic formulation of numerical relativity \cite{w12} require its adaptability and extension to a variety of scenarios. 
Although the Cauchy approach
in numerical relativity has proven relatively successful
in the simulation of binary black holes \cite{puncture},
the accurate prediction of wave forms from black
hole-black hole, black hole-neutron star, black hole-boson
star binaries as sources of gravitational radiation stands
as formidable pending problems. 
The characteristic formulation of general relativity
offers an alternative
for the accurate prediction of waveforms from such
astrophysical scenarios, but further improvements
are mandatory to make it attractive and competitive.}

{One of the prime factors affecting the accuracy of any
characteristic code is the introduction of a smooth coordinate 
system covering the sphere, which labels the null
directions on the outgoing (ingoing) light cones. Interestingly, 
this is also an underlying problem in meteorology
and oceanography \cite{p57}. The LEO code, a large scale computational
framework based on the characteristic formulation \cite{gbf07}, was inspired 
by global forecasting techniques \cite{s72}, \cite{rip96} and showed
great potential in handling 3D problems. 
The solver was tested solving the Einstein-Klein-Gordon (EKG). Despite its simplicity,
analytical studies of this toy model for a self-gravitating massless scalar field 
show that it exhibits highly nonlinear physics \cite{christo,hs,koikeetal,LRR}.}
One dimensional numerical simulations of the EKG led to the discovery of critical phenomena \cite{c93} and {to reveal some} features of the asymptotic
spacetime structure. {For instance, the
Bondi mass and the scalar monopole moment satisfy an asymptotic relation at high amplitudes \cite{gw92}. The Bondi mass and news function reflect the discretely self-similar behavior \cite{pha05}.}

{Extending the work of {G\'omez et al. \cite{gpw94}} and Papadopoulos \cite{p02},
here we incorporate a massless scalar
field and solve the 2D EKG system. 
We perform numerical validations that include tests of convergence, the simulation of the exponentially damped oscillation modes, called quasinormal modes (QNM) and the energy conservation (in the linear approximation) for the massless scalar field. Gravitational radiation waveforms and the nonlinear regime deserve especial attention and are postponed for a future study. However, we include a calculation of the distorted marginally trapped two-surface. The solver developed can be
considered as an intermediate step, both in computational cost and
dimensions. The 2D EKG is an interesting problem in itself, well suited to explore global issues \cite{pha05}, \cite{b14}.}
\begin{figure}
\includegraphics[width=2.in,height=2.5in,angle=0]{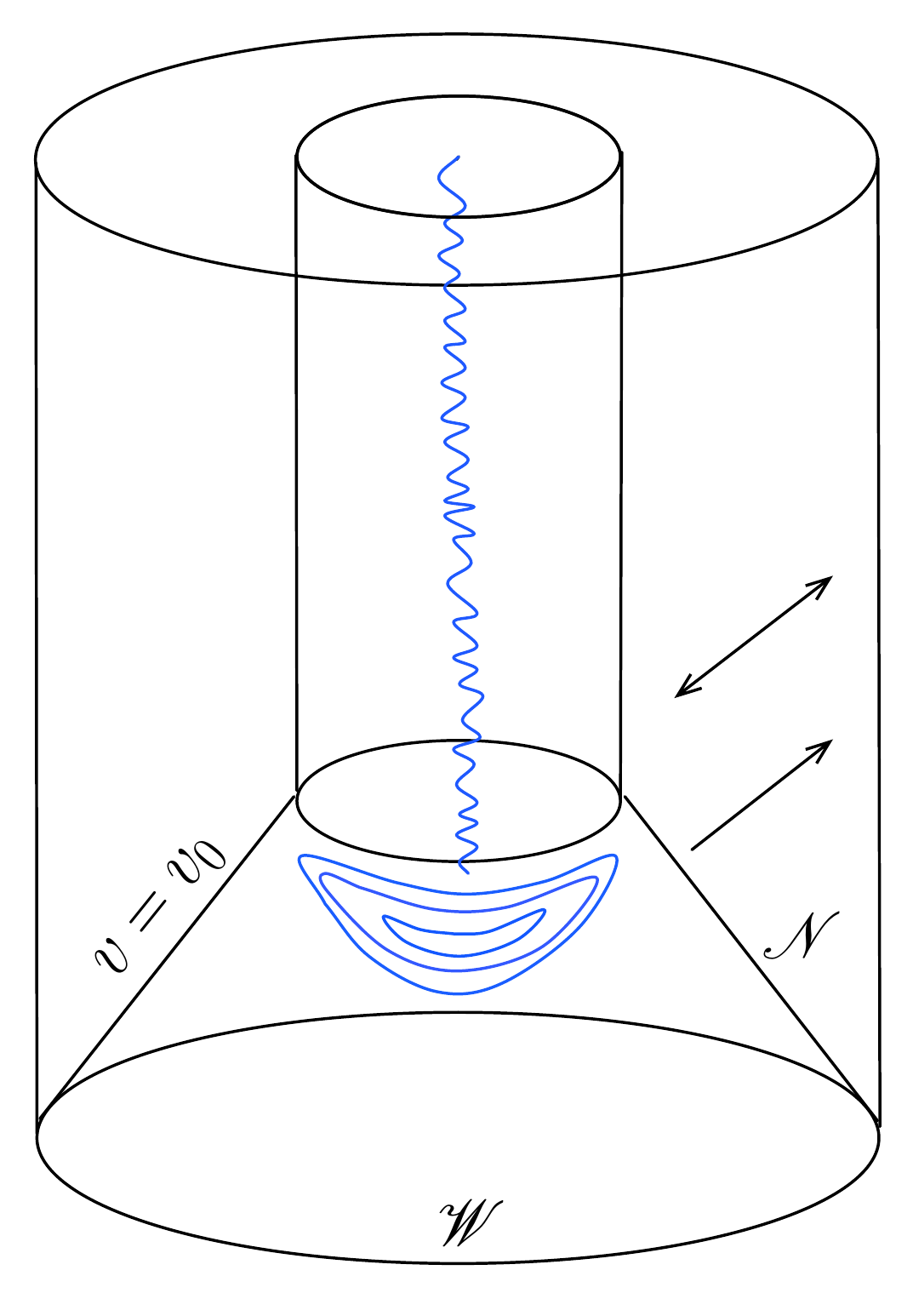}
\caption{Spacetime diagram with the problem setup. The foliation is based on advanced time $v$; the geometry of the world tube ($\mathscr{W}$) is kept fixed at all times and is given by the Schwarzschild values. The ingoing light cones emanate from $\mathscr{W}$. The initial ingoing light cone $\mathscr{N}$ at $v_0$ is distorted with the specification of an arbitrary outgoing massless scalar field. In general, the evolution generates gravitational radiation that, together with scalar radiation, is scattered toward and away the distorted black hole.}
\label{fig:ingoing_set_up}
\end{figure}

{The paper is organized as follows. In Sec. II we setup the EKG system for a massless scalar field minimally coupled to gravity. In this section we also consider issues about QNM, energy conservation and marginally trapped surfaces. Sec. III is dedicated to details about the numerical implementation, and to the tests of convergence to second order of the nonlinear solver. In Sec. IV we present our results. Finally we conclude in Sec. V with some remarks.}
\section{Setup}
\subsection{The EKG problem}
In general, the field equations for a massless scalar field minimally coupled to gravity are
\begin{equation}
G_{ab}=-8\pi T_{ab},
\end{equation}
with 
\begin{equation}
T_{ab}=\nabla_a\Phi\nabla_b\Phi-\frac{1}{2}g_{ab}\nabla^c\Phi\nabla_c\Phi,
\end{equation}
and can be reduced to
\begin{equation}
R_{ab}=-8\pi\nabla_a\Phi\nabla_b\Phi,
\end{equation}
which have to be considered together with the wave equation
\begin{equation}
\square\Phi=0,
\end{equation}
in order to complete the EKG system. 
\subsection{Two--dimensional ingoing formulation}
The initial-boundary value problem is formulated following the Winicour-Tamburino framework \cite{wt65,tw66}, with ingoing light cones emanating from a timelike world tube $\mathscr{W}$ (see Fig. \ref{fig:ingoing_set_up}). The characteristic initial value problem of the EKG system can be explicitly written using the ingoing line element in the case of axial and reflection symmetry \cite{bvm62}
\begin{eqnarray}
ds^2&=&\bigg(\frac{V}{r}e^{2\beta}-U^2r^2e^{2\gamma}\bigg)dv^2-2e^{2\beta}dvdr\nonumber\\
&-&2Ur^2e^{2\gamma}dvd\theta-r^2(e^{2\gamma}d\theta^2+e^{-2\gamma}\sin^2\theta d\phi^2),\nonumber\\ \label{ingoing_metric}
\end{eqnarray}
where $\beta=\beta(v,r,\theta)$, $V=V(v,r,\theta)$, $U=U(v,r,\theta)$, $\gamma=\gamma(v,r,\theta).$ This metric is twist-free, and therefore rotation is not permitted. 
Thus we get the following field hypersurface equations
\begin{equation}
\beta_{,r}=\frac{1}{2}{r}\left(\gamma_{,r}^2 + \psi_{,r}^2\right), \label{e1}
\end{equation}
\begin{eqnarray}
(r^2\mathcal{Q})_{,r}&=&2r^2\left[2(\gamma_{,r}\gamma_{,\theta}+\psi_{,r}\psi_{,\theta})+r^2\left(\frac{\beta}{r^2}\right)_{,r\theta}\right.\nonumber\\
&-&\left. \frac{1}{\sin^2\theta}(\sin^2\theta\gamma)_{,r\theta}\right],\label{e2}
\end{eqnarray}
\begin{equation}
U_{,r}=e^{2(\beta-\gamma)}\frac{\mathcal{Q}}{r^2},\label{e3}
\end{equation}
\begin{eqnarray}
V_{,r}&=&e^{2(\beta-\gamma)}\bigg[1+(3\gamma_{,\theta}-\beta_{,\theta})\cot\theta+\gamma_{,\theta\theta}\nonumber\\
&-&\psi_{,\theta}^2-\beta_{,\theta\theta}-\beta_{,\theta}^2-2\gamma_{,\theta}
(\gamma_{,\theta}-\beta_{,\theta})\bigg]\nonumber\\
&+&\frac{r}{2\sin\theta}\bigg[\sin\theta(rU_{,r}
+4U)\bigg]_{,\theta}\nonumber\\
&-&\frac{1}{4}e^{2(\gamma-\beta)}r^4U_{,r}^2,
\label{e4}
\end{eqnarray}
and the evolution equations
\begin{eqnarray}
&-&e^{2\beta}\square^{(2)}_+(r\gamma)=-\left(\frac{V}{r}\right)_{,r}\gamma +\frac{1}{4}r^3e^{2(\gamma-\beta)}U_{,r}^2 \nonumber\\
&+&\frac{1}{r}e^{2(\beta-\gamma)}(\psi^2_{,\theta}+\beta_{,\theta\theta}+\beta_{,\theta}^2-\beta_{,\theta}\cot\theta)\nonumber\\
&-&\frac{1}{2r}[r^2(2\gamma_{,\theta}U+U_{,\theta}-U\cot\theta)]_{,r}\nonumber\\
&-&\frac{r}{\sin\theta}(\gamma_{,r}U\sin\theta)_{,\theta},
\label{e5}
\end{eqnarray}
\begin{eqnarray}
&-&e^{2\beta}\square^{(2)}_+(r\psi)=-\left(\frac{V}{r}\right)_{,r}\psi 
-\frac{1}{r}(r^2\psi_{,\theta}U)_{,r}\nonumber\\
&+&\frac{1}{r\sin\theta}[\sin\theta(e^{2(\beta-\gamma)}\psi_{,\theta}-r^2U\psi_{,r})]_{,\theta},\label{e6}
\end{eqnarray}
where $\psi=2\sqrt{\pi}\Phi$ and $\square^{(2)}_+f=e^{-2\beta}[2f_{,vr}{+}(Vf_{,r}/r)_{,r}]$.

\subsection{Scattering off a Schwarszhild black hole}
On $\mathscr{W}$ we can specify the boundary conditions in many ways. In this work we do the following choice: 
\begin{subequations}
\begin{eqnarray}
\beta(v,r=r_{\mathscr{W}},\theta)&=&0,\\
U(v,r=r_{\mathscr{W}},\theta)&=&0,\\
V(v,r=r_{\mathscr{W}},\theta)&=&r-2M, 
\end{eqnarray}
\label{bc}
\end{subequations}
where $M$ is the black hole mass.
The geometry of $\mathscr{W}$ is kept fixed at all times and is given by the Schwarzschild values. To be consistent we set the initial conditions
\begin{equation}\gamma(v=v_0,r,\theta)=0\label{icgamma}\end{equation} 
and
\begin{equation}
\psi(v=v_0,r,\theta)=\frac{\lambda}{r} e^{-(r-r_0)^2/\sigma^2}P_\ell(\theta), \label{icpsi}
\end{equation}
where $\sigma=0.5M$, $r_0=3M$ and $P_\ell(\theta)$ are the Legendre polynomials. 
In this way the initial ingoing light cone $\mathscr{N}$ at $v_0$ is distorted with the specification of an arbitrary outgoing massless scalar field. In general, with the nonlinear evolution, the black hole is distorted and the gravitational radiation is generated.

\subsection{Scalar field on a fixed background}
The above system of equations (\ref{e1})--(\ref{e6}) describes a self--gravitating scalar field. In the limit of small amplitudes, $|\psi|\ll 1$, the scalar field can be treated as a perturbation propagating on a fixed background. This considerably simpler model is contained in the fully nonlinear case, and is implemented in our code by integrating only Eq. (\ref{e6}). For a Schwarzschild background, the metric reads
\begin{equation}
ds^2=(1-2M/r)dv^2-2dvdr-r^2(d\theta^2+\sin^2\theta d\phi^2).
\end{equation}
On this fixed background the linear approximation of Eq. (\ref{e6}) is
\begin{eqnarray}
2(r\psi)_{,vr}+[(1-2M/r)(r\psi)_{,r}]_{,r}=\nonumber\\
\frac{2M\psi}{r^2} - \frac{1}{r\sin\theta}[\sin\theta \psi_{,\theta}]_{,\theta}.
\label{e6_linear_a}
\end{eqnarray}
For the simulations in the present work, we will be interested in solutions of the 2D scalar field on a fixed background, as discussed in the next subsection.
\subsection{QNM in a Schwarzschild background}
The linear equation for the scalar field on a fixed background, Eq. (\ref{e6}), is separable; i.e. its solutions can be written in the form
\begin{equation}
\psi(v,r,\theta)=\frac{1}{r}\sum_{\ell=0}^{\infty}\chi_\ell(v,r)P_\ell(\theta)
\end{equation}
 where each of the $\chi_\ell$ satisfies the one--dimensional wave equation in the plane $(v, r)$,
\begin{equation}
2\chi_{,vr}+[(1-2M/r)\chi_{,r}]_{,r}=\left[\frac{2M}{r^3}+\frac{\ell(\ell+1)}{r^2}\right]\chi. \label{e6_linear_b}
\end{equation}
This last equation is the usual which govern the scalar perturbations of a Schwarzschild black hole \cite{n99}, written here in characteristic coordinates $(v, r, \theta)$. Equation (\ref{e6_linear_b}) has been studied extensively \cite{n99,ns92,k04}, its most salient feature being the existence of QNM, whose frequencies have been tabulated; see for example Ref. \cite{k04}. In the present work we will use both the QNM equation, Eq. (\ref{e6_linear_b}), and the linear Eq. (\ref{e6_linear_a}), as tests to validate our numerical implementation. We do this in an incremental fashion, solving Eq. (\ref{e6_linear_b}) for fixed values of $\ell$, and comparing the effectiveness of the numerical integration scheme and of our boundary conditions in reproducing the QNM. To this end, we implement a purely radial code for Eq. (\ref{e6_linear_b}) that employs the same numerical integration scheme that is used in the ``linear'' code [which solves Eqs. (\ref{e6_linear_a}) and (\ref{e6_linear_b})], and in the full nonlinear code. 
In ingoing null coordinates, the slices at $v=$ const. penetrate the event horizon 
at $r=2M$, effectively providing for an excision scheme, where the evolution can be stopped at a finite number of points inside the boundary, because the behavior of the field inside the horizon does not affect the solution outside. Evolutions in ingoing coordinates are carried out on a radial grid, for which boundary data are required at a fixed value of $r_{\text{out}} > 2M$. Because of the presence of this outer boundary, simulations in ingoing coordinates can only be run for a limited time, typically $v \sim 2r_{\text{out}}$, before outer boundary effects influence the signal extracted.

{\subsection{Energy carried out by the scalar field}
We calculate the balance of the scalar field energy contained between the inner and the outer boundary \cite{gbf07}. The expressions we give here are valid in the linear case, where the background metric is the Schwarzschild metric. For a more general approach to this issue, the linkage integrals have to be calculated; specifically the asymptotic Killing vector field must be parallelly propagated from null infinity \cite{wt65}.
Restricted to the background case, then, given a Killing vector field $\xi^a$ of the metric $g_{ab}$, $\pounds_\xi g_{ab}=0$, we can define the conserved quantity
\begin{equation}
{\cal C}=\int \xi^a T^b_a  d\Sigma_b.
\label{linkage}
\end{equation}
In particular, selecting the timelike Killing vector $\xi^a=\delta^a_v$, and for a surface of constant $v$, ${\cal C}$ is the energy contained on the surface,
\begin{equation}
E(v)=\int T^v_v dV,
\end{equation}
where $dV$ is the volume element of the surface at constant
$v$. For a sphere at constant $r$, $\cal{C}$ represents the energy flux across the surface,
\begin{equation}
P(v)=\int T^r_v r^2 d\Omega,
\end{equation}
with $d\Omega$ the solid angle element. The relevant components the energy-momentum tensor a massless scalar field are
\begin{equation}
T^v_{\;\;v}=e^{-2\beta} \psi_{,r}\left[\frac{V}{r}\psi_{,r}-2U\psi_{,\theta}\right]+ \frac{2}{r^2}\psi^2_{,\theta}{e^{-2\gamma}}
\end{equation}
\begin{equation}
T^r_{\;\;v}={-2e^{-2\beta}\psi_{,v}\left[\frac{V}{r}\psi_{,r}+\psi_{,v}-U\psi_{,\theta}\right]}
\end{equation}
In the case of a linear scalar perturbation on a Schwarzschild background, the energy content of a hypersurface at constant $v$ is given by
\begin{equation}
E(v)=2\pi \int \left[\left(1-\frac{2M}{r}\right)(r\psi_{,r})^2+2\psi^2_{,\theta}\right]dr\sin\theta d\theta,
\label{ten}
\end{equation}
and the power radiated at time $v$ across a surface of constant $r$ is
\begin{equation}
P=-4\pi r^2 \int \psi_{,v}\left[\psi_{,v}+\left(1-\frac{2M}{r}\right)\psi_{,r}\right]\sin\theta d\theta.
\label{power}
\end{equation}
In our simulations we place $r_{\text{in}}$ close enough to the Schwarzschild black hole, and the outer boundary such as  $2M< r_{\text{out}} < r_{\mathscr{W}}$. 
For the flux across the inner (outer) boundary, the integral as well as the spatial and time derivatives are to be taken as evaluated at 
$r=r_{\text{in}} (r_{\text{out}})$. With these definitions, the following energy conservation law holds,
\begin{equation}
\Sigma(v)=E(v)+ \int^v_{v_0}[P_{\text{in}}(v')-P_{\text{out}}(v')]dv'=\text{const}.
\label{conser}
\end{equation}
The expressions given above hold only in the limit in which $\partial_t$ is a Killing vector of the metric, so we use them as a criterion for code testing.}

{\subsection{Marginally trapped surfaces}
In general, the constructed spacetime by the present approach contains a distorted black hole. This is made geometrically precise by the introduction of the concept of a marginally trapped two-surface (MTS)  on a given ingoing light cone $\mathcal{N}$.
A MTS is defined, in this context, as the two-parameter radial function $\mathcal{R} ( r , \theta )$ on which the expansion $\Theta_l$ of an outgoing null ray pencil $l^\alpha$ vanishes \cite{gmw97}.  
{If $n^\alpha$ is tangent to the generators of $\mathcal{N}$, we get
\begin{equation}
n_\alpha=g_{rv}v_{,\alpha}.
\end{equation}
\noindent Thus, for the diverging slices $\mathcal{S}$ of $\mathcal{N}$, given
by $\mathcal{R}=r-R(\theta)$, the outgoing normal $l_\alpha$ to $\mathcal{S}$ is 
\begin{eqnarray}
l_\alpha&=&{r}_{,\alpha}-\theta_{,\alpha} R_{,\theta} \nonumber\\
&-&\frac{1}{2}g_{rv}\left[g^{rr}+g^{\theta\theta}R^2_{,\theta}-2g^{r\theta}R_{,\theta}\right]v_{,\alpha}.
\end{eqnarray}
For the projection tensor $h_{\alpha\beta}$ into the tangent space of $\mathcal{S}$
\begin{equation}
h^\alpha_\beta=g^\alpha_\beta-n^\alpha l_\beta -l^\alpha n_\beta,
\end{equation}
the expansion associated to the null vector $l^\alpha$ can be written as
\begin{eqnarray}
\Theta_l&=&2h^{\alpha\beta}\nabla_\alpha l_\beta,
\end{eqnarray}
which explicitly is
\begin{widetext}
\begin{eqnarray}
\frac{1}{2}e^{2\gamma}r^2\Theta_l&=&
R_{,\theta} \left[\cot\theta+2(\beta_{,\theta}-\gamma_{,\theta})+U_{,r}e^{2(\gamma-\beta)}\right]+R^2_{,\theta}\left[2(\beta_{,\theta}-\gamma_{,\theta})-\frac{1}{r}\right]+R_{,\theta\theta} \nonumber\\
&+&r^2\left[U_{,\theta}+U\cot\theta-\frac{V}{r^2}\right]e^{2(\gamma-\beta)}.
\label{expansion}
\end{eqnarray}
\end{widetext}
This is an elliptic equation and has to be solved numerically with the system evolution. The convergence to $\Theta_l=0$ leads to $R(\theta)$, which locates the MTS.}

{As an indicator of the trapped horizon location we can estimate the MTS using the q-boundary, following the method {detailed} in Ref. \cite{gmw97}. 
{If $R=$ const., Eq. (\ref{expansion}) reduces to
\begin{equation}
q\equiv\frac{1}{2}e^{2\beta}r^2\Theta_l=r^2(U_{,\theta}+U\cot\theta)-V. \label{Q}
\end{equation}
The q-boundary is the slice with the largest $r=$ const. on which $q\le 0$. Such slice has $\Theta_l<0$, therefore is trapped, and trapped surfaces are inside the MTS. In the nonvacuum spherical symmetric case the MTS is given by $V=0$, which determines the location of the apparent horizon at $r=2M$ (corresponding in position with the event horizon for the vacuum Schwarzschild metric). In the absence of spherical symmetry $\Theta_l$ vanishes at points for which $q=0$. Thus, the q-boundary is everywhere trapped or marginally trapped and is a simple algebraic procedure for locating an inner boundary inside an event horizon.}}

%%%%%%%%%%%%%%%%%%%%%%%%%%%%%%%%%%%
\section{Numerical implementation}
%%%%%%%%%%%%%%%%%%%%%%%%%%%%%%%%%%%
The computational algorithm is related to those developed in Refs. \cite{gpw94}, \cite{p02} and, as we shall see, was shown to be second order accurate in the nonlinear regime. In the linear regime we get the expected QNM and the energy conservation. We briefly review some issues about the regularization and the discretization of equations.
\subsection{Regularization}
The coordinate system consists of a radial $r$ coordinate and an angular $y=-\cos\theta$ coordinate. The numerical grid is uniformly spaced in both coordinates. Also we use the regularized variables $\hat\gamma=\gamma/\sin^2\theta$,
$\hat U={U}/{\sin\theta}$, $\mathcal{\hat Q}={\mathcal{Q}}/{\sin\theta}$. Thus, the hypersurface equations are given by
\begin{equation}
\beta_{,r}=\frac{r}{2}\left(\hat{\gamma}_{,r}^2(1-y^2)^2 + \psi^2_{,r}\right), \label{betar}
\end{equation}
\begin{eqnarray}
(r^2\mathcal{\hat Q})_{,r}&=&2r^2\left\{2(1-y^2)\hat\gamma_{,r}[(1-y^2)\hat\gamma_{,y}-2y\hat{\gamma}]\right.\nonumber\\
&+&2\psi_{,r}\psi_{,y}+r^2\left({\beta}/{r^2}\right)_{,ry}
+4y\hat{\gamma}_{,r}\nonumber\\
&-&\left.(1-y^2)\hat{\gamma}_{,ry}\right\},
\label{Qhat}
\end{eqnarray}
\begin{equation}
\hat{U}_{,r}=e^{2[\beta-\hat\gamma(1-y^2)]}\frac{\mathcal{\hat Q}}{r^2}
\label{Uhat}
\end{equation}
{\begin{eqnarray}
V_{,r}&=&e^{2[\beta-\hat\gamma(1-y^2)]}\bigg\{1+(1-y^2)^2\hat\gamma_{,yy}-2(1-5y^2)\hat\gamma\nonumber\\
&-&8y(1-y^2)\hat\gamma_{,y}+2y\beta_{,y}-(1-y^2)(\beta_{,yy}+\beta_{,y}^2) \nonumber\\
&-&(1-y^2)\psi^2_{,y}-2(1-y^2)[\hat \gamma_{,y}(1-y^2)\nonumber\\
&-&2y\hat \gamma]^2+2(1-y^2)(\hat \gamma_{,y}(1-y^2)-2y\hat \gamma)\beta_{,y}\bigg\}\nonumber\\
&+&\frac{r}{2}\bigg[(1-y^2)(r\hat U_{,ry}+4\hat U_{,y})-2yr\hat U_{,r}-8y\hat U\bigg]
\nonumber\\
&-&\frac{1}{4}e^{2(\hat\gamma(1-y^2)-\beta)}r^4\hat U_{,r}^2(1-y^2)  \label{Ver}
\end{eqnarray}}
and the evolution equations 
{\begin{eqnarray}
-e^{2\beta}\square^{(2)}_+(r\hat\gamma)=&-&\left(\frac{V}{r}\right)_{,r}\hat\gamma +\frac{1}{4}r^3e^{2(\hat\gamma(1-y^2)-\beta)}\hat U_{,r}^2 \nonumber\\
&+&\frac{1}{r}e^{2(\beta-\hat\gamma(1-y^2))}(\psi^2_{,y}+\beta_{,yy}+\beta_{,y}^2)\nonumber\\
&-&r\bigg[(\hat\gamma_{,ry}\hat U+\hat\gamma_{,r}\hat U_{,y})(1-y^2)\nonumber\\
&-&4y\hat\gamma_{,r}\hat U\bigg] - (\hat\gamma_{,y}(1-y^2)-2y\hat\gamma)2\hat U\nonumber\\
&-&\hat U_{,y}-\frac{r}{2}\bigg[(\hat\gamma_{,yr}(1-y^2)-2y\hat\gamma_{,r})2\hat U\nonumber\\
&+&(\hat\gamma_{,y}(1-y^2)-2y\hat\gamma)2\hat U_{,r}\nonumber\\
&+&\hat U_{,yr})\bigg], \label{gammaevol}
\end{eqnarray}}
{\begin{eqnarray}
-e^{2\beta}\square^{(2)}_+(r\psi)=&-&\left(\frac{V}{r}\right)_{,r}\psi \nonumber\\
&-&2(1-y^2)\psi_{,y}\hat U\nonumber\\
&-&r(1-y^2)[\psi_{,ry}\hat U+\psi_{,y}\hat U_{,r}]\nonumber\\
&+&\frac{1}{r}\bigg\{(1-y^2)\big\{2[\beta_{,y}-\hat\gamma_{,y}(1-y^2)\nonumber\\
&+&2y\hat\gamma]\psi_{,y}e^{2[\beta-\hat\gamma(1-y^2)]}\nonumber\\
&+& e^{2[\beta-\hat\gamma(1-y^2)]} \psi_{,yy}\nonumber\\
&-&r^2(\hat U_{,y}\psi_{,r}+\hat U\psi_{,ry})\big\}\nonumber\\
&-&2y(\psi_{,y}e^{2[\beta-\hat\gamma(1-y^2)]}\nonumber\\
&-&r^2\hat U\psi_{,r})\bigg\}.
\label{psievol}
\end{eqnarray}}
We now review the essentials of the numerical integration procedure. 
\subsection{Discretization}
\subsubsection{Hypersurface equations}
Equation (\ref{betar}) is easily discretized to get at once
\begin{equation}
\beta^n_{j,i}=\beta^n_{j,i-1} - \Delta r \beta_{,r}|^n_{j,i-1/2},
\end{equation}
where indexes $n$, $j$ and $i$ indicate discretization in $v$, $\theta$ and $r$, respectively.
In general, any first order radial derivative is calculated as
$f_{,r}|_{i-1/2}=-{(f_i-f_{i-1})}/{\Delta r}$, because we proceed from $r_{\mathscr{W}}$ to $r_{\mathscr{B}}$ ($r_{\mathscr{W}}> r_{\mathscr{B}}$), where $r_\mathscr{B}$ is the inner boundary,
$\Delta r={(r_{\mathscr{W}}-r_{\mathscr{B}})}/{N_r}$,
and $N_r$ the number of grid points in $r$. Thus, the term $\beta_{,r}|^n_{j,i-1/2}$ is evaluated as
\begin{equation}
\beta_{,r}|^n_{j,i-1/2}=\frac{1}{2}r_{i-1/2}(\psi^2_{,r}+\hat\gamma^2_{,r})|_{j,i-1/2},
\end{equation}
where derivatives of the RHS are evaluated numerically, as indicated above.

{
Now, combining Eqs. (\ref{Qhat}) and (\ref{Uhat}) we get
\begin{equation}
\mathcal{C}\hat U_{,r}+\frac{1}{2}r^2\hat U_{,rr}=\mathcal{H}_{\hat U}e^{2(\beta-\gamma)},
\label{UE}
\end{equation}
where 
$$\mathcal{C}=r^2\left[\frac{2}{r}+\hat\gamma_{,r}(1-y^2)-\beta_{,r}\right],$$
\begin{eqnarray}
\mathcal{H}_{\hat U}=2(1-y^2)\hat\gamma_{,r}[(1-y^2)\hat\gamma_{,y}-2y\hat{\gamma}]+\bar\psi_{,r}\psi_{,y}+\bar\psi_{,y}\psi_{,r}\nonumber\\
+r^2\left(\frac{\beta}{r^2}\right)_{,ry}
+4y\hat{\gamma}_{,r}-(1-y^2)\hat{\gamma}_{,ry}. \nonumber
\end{eqnarray}
Using centered finite differences at $j$, $i-1$, for stability reasons previously established in Ref. \cite{gpw94}, and dictated by the second order derivative in $r$,
we discretized Eq. (\ref{UE}); any other discretization in this same equation is staggered at $j+1/2$, $i-1/2$. Thus, we get
\begin{eqnarray}
&&-\frac{\mathcal{C}^n_{j+1/2,i-1/2}}{2\Delta r}(\hat U^n_{j,i}-\hat U^n_{j,i-2}) \nonumber \\
&&+ \frac{r^2_{i-1/2}}{2\Delta r^2}(\hat U^n_{j,i} -2\hat U^n_{j,i-1} + \hat U^n_{j,i-2})\nonumber\\
&&=[\mathcal{H}_{\hat U}e^{2(\beta-\gamma)}]^n_{j+1/2,i-1/2},
\end{eqnarray}
from which we obtain
\begin{eqnarray}
\hat U^n_{j,i}&=&\frac{1}{f_a}\left\{ \Delta r^2 [\mathcal{H}_{\hat U}e^{2[\beta-\hat\gamma(1-y^2)]}]^n_{j+1/2,i-1/2} \right.\nonumber\\
&-&\left.(f_b\hat U^n_{j,i-1}+f_c \hat U^n_{j,i-2})\right\},
\end{eqnarray}
where
$$f_a=\frac{1}{2}(r_{i-1/2}^2-\Delta r \mathcal{C}^n_{j+1/2,i-1/2}),$$
$$f_b=-r_{i-1/2}^2,$$
$$f_c=\frac{1}{2}(r_{i-1/2}^2+\Delta r \mathcal{C}^n_{j+1/2,i-1/2}).$$}

{Next, to discretize Eq. (\ref{Ver}) we define the mass aspect
\begin{equation}
M(v,r,\theta)=\frac{1}{2}(r-V).
\end{equation}
which leads to
\begin{equation}
M_{,r}=\frac{1}{2}(1-\mathcal{H}_{M\hat U}-\mathcal{H}_{M\beta\hat\gamma\psi}),
\end{equation}
where
\begin{eqnarray}
\mathcal{H}_{M\hat U}&=&\frac{r}{2}\bigg[(1-y^2)(r\hat U_{,ry}+4\hat U_{,y})-2yr\hat U_{,r}-8y\hat U\bigg]\nonumber\\
&-&\frac{1}{4}e^{2(\hat\gamma(1-y^2)-\beta)}r^4\hat U_{,r}^2(1-y^2),
\end{eqnarray}
\begin{eqnarray}
\mathcal{H}_{M\beta\hat\gamma\psi}&=&e^{2[\beta-\hat\gamma(1-y^2)]}\bigg\{1+(1-y^2)^2\hat\gamma_{,yy}-2(1-5y^2)\hat\gamma\nonumber\\
&-&8y(1-y^2)\hat\gamma_{,y}+2y\beta_{,y}-(1-y^2)(\beta_{,yy}+\beta_{,y}^2)
\nonumber\\
&-&(1-y^2)\psi^2_{,y}-2(1-y^2)[\hat \gamma_{,y}(1-y^2)\nonumber\\
&-&2y\hat \gamma]^2+2(1-y^2)(\hat \gamma_{,y}(1-y^2)\nonumber\\
&-&2y\hat \gamma)\beta_{,y}\bigg\}. 
\end{eqnarray}
In this way we obtain
\begin{eqnarray}
M^n_{j,i}&=&M^n_{j,i-1}\nonumber\\
&-&\Delta r(1-\mathcal{H}_{M \hat U}|^n_{j-1/2,i-1} -
\mathcal{H}_{M \beta\hat\gamma\psi}|^n_{j,i-1/2}).\nonumber\\
&& \, 
\end{eqnarray}
Observe that the discretization is staggered (backward) for $\hat U$ terms
with respect to the other terms.}

All the obtained formulae for the hypersurface equations are recursive and can be applied from $r\leq r_{\mathscr{W}}$ to $r_{\mathscr{B}}$. 
%%%%%%%%%%%%%%%%%%%%%%%%%%%%%%%%%%%%
\subsubsection{Evolution equations}
%%%%%%%%%%%%%%%%%%%%%%%%%%%%%%%%%%%%
\begin{figure}
\includegraphics[width=3.5in,height=3.5in,angle=0]{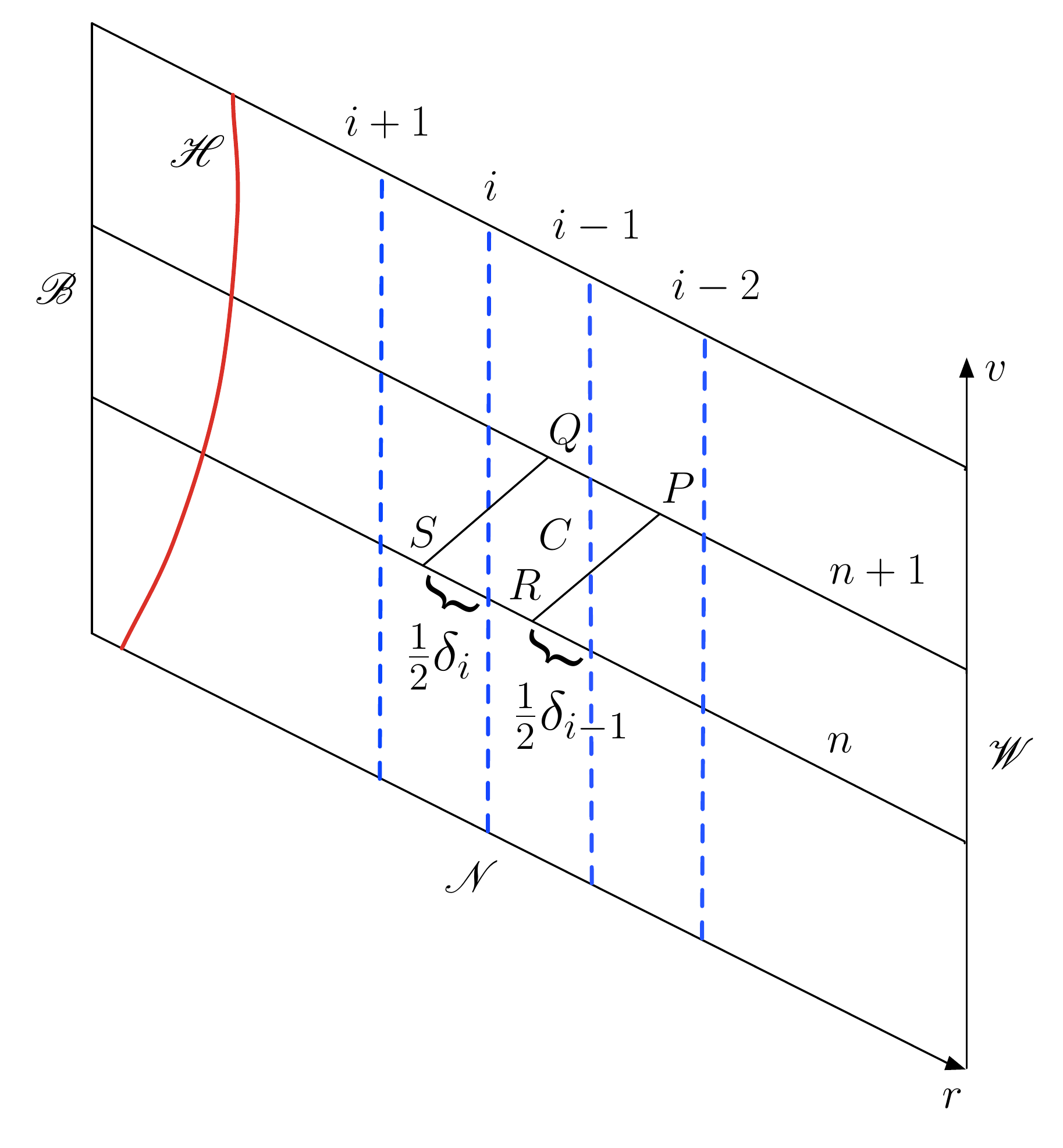}
\caption{Ingoing marching algorithm for one particular angle. Initial data are set  on the initial null cone $\mathscr{N}$; boundary conditions are set on the world tube $\mathscr{W}$. Successive levels of the advanced time $v$ are depicted (diagonal lines $n$ and $n+1$) along with the radial grid (dashed vertical lines $i-2$, $i-1$, $i$, $i+1$). The radial grid starts at $\mathscr{W}$ ($i=0$) and terminates inside the horizon $\mathscr{H}$, at the inner boundary $\mathscr{B}$ ($i=N_r$). The evolution equation for $\hat \gamma$ relates the field values $\hat \gamma_R$, $\hat \gamma_S$, $\hat \gamma_P$, $\hat \gamma_Q$ to the value of $\cal{H}_C$. The same structure of the integration procedure applies to $\psi$.}
\label{fig:2}
\end{figure}
{The discretization of the evolution equation (\ref{gammaevol}) proceeds in detail as follows (see Refs. \cite{gw92}, \cite{gpw94}, \cite{p02}).
The core of the evolution integration is the ingoing marching algorithm from $r_{\mathscr{W}}$ to $r_{\mathscr{B}}$ involving the two time levels $n$ and $n+1$.

We integrate Eq. (\ref{gammaevol}) over the null parallelogram $\mathcal{A}$ formed by  ingoing and outgoing radial null rays in the $(v,r)$ plane that intersect at vertices $P$, $Q$, $R$ and $S$, as depicted in Fig. \ref{fig:2}. Thus, we have
\begin{equation}
\int_\mathcal{A} e^{2\beta} \square^{(2)}_+\tilde\gamma dvdr=\int_\mathcal{A} \mathcal{H} dvdr,
\end{equation}
where $\tilde\gamma=r\hat\gamma$ and $\mathcal{H}$ is the RHS of Eq. (\ref{gammaevol})
with the changed sign. Using the mean value theorem, we approximate
\begin{equation}
\int_\mathcal{A} \mathcal{H} dvdr=\mathcal{H}_C\int_\mathcal{A} drdv,
\end{equation}
where the subscript $C$ indicates that the quantity is evaluated at the center
of the null parallelogram $\Sigma$. Now, easily we can get exactly
\begin{equation}
\int_\mathcal{A} drdv=\frac{1}{2} \Delta v (r_Q-r_P+r_S-r_R).
\end{equation}
On the other hand, we use the conformal invariance \cite{gw92} to get 
\begin{equation}
\int_\mathcal{A} e^{2\beta} \square^{(2)}_+\tilde\gamma dvdr=2(\tilde\gamma_Q-\tilde\gamma_P+\tilde\gamma_R-\tilde\gamma_S).
\end{equation}
Thus, the marching algorithm reads
\begin{equation}
\tilde\gamma_Q=\tilde\gamma_P+\tilde\gamma_S-\tilde\gamma_R + \frac{1}{4}\Delta v (r_Q-r_P+r_S-r_R)\mathcal{H}_C. \label{marching_alg}
\end{equation}

The numerical implementation of this formula proceeds as follow. Referring again to Fig. \ref{fig:2}, interpolations are required, and can be linear to keep a globally second order approximation.}
{The ingoing null geodesic equation is given by
\begin{equation}
\frac{dr}{dv}=\frac{1}{2}\left(1-\frac{2M}{r}\right).
\end{equation}
Thus, the displacements $\delta_i$ and $\delta_{i-1}$ are calculate as
\begin{equation}
\delta_i=\frac{1}{2}\left[1-\left(\frac{M^{n-1}_{j,i+1}}{r_{i+1}}+\frac{M^{n}_{j,i-1}}{r_{i-1}}\right)\right]\Delta v
\end{equation}
and
\begin{equation}
\delta_{i-1}=\frac{1}{2}\left[1-\left(\frac{M^{n-1}_{j,i}}{r_{i}}+\frac{M^{n}_{j,i-2}}{r_{i-2}}\right)\right]\Delta v.
\end{equation}
The vertices coordinates are positioned by
\begin{subequations}
\begin{eqnarray}
r_P&=&r_{i-1} + \frac{1}{2}\delta_{i-1},\\
r_R&=&r_{i-1} - \frac{1}{2}\delta_{i-1},\\
r_Q&=&r_i + \frac{1}{2}\delta_{i},\\
r_S&=&r_i - \frac{1}{2}\delta_{i},
\end{eqnarray}
\end{subequations}
and the center coordinate by
\begin{equation}r_C=\frac{1}{2}(r_P+r_S).\end{equation}
The interpolate field at each corner of the null parallelogram is
\begin{subequations}
\begin{eqnarray}
\tilde\gamma_{j,P}&=&\tilde\gamma^{n}_{j,i-1}-\frac{1}{2}\frac{\delta_{i-1}}{\Delta r}(\tilde\gamma^{n}_{j,i-1}-\tilde\gamma^{n}_{j,i-2}),\\
\tilde\gamma_{j,R}&=&\tilde\gamma^{n-1}_{j,i-1}+\frac{1}{2}\frac{\delta_{i-1}}{\Delta r}(\tilde\gamma^{n-1}_{j,i}-\tilde\gamma^{n-1}_{j,i-1}),\\
\tilde\gamma_{j,S}&=&\tilde\gamma^{n-1}_{j,i}+\frac{1}{2}\frac{\delta_{i}}{\Delta r}(\tilde\gamma^{n-1}_{j,i+1}-\tilde\gamma^{n-1}_{j,i}),\\
\tilde\gamma_{j,Q}&=&\tilde\gamma^{n}_{j,i}-\frac{1}{2}\frac{\delta_{i}}{\Delta r}(\tilde\gamma^{n}_{j,i}-\tilde\gamma^{n}_{j,i-1}).
\end{eqnarray}
\end{subequations}
To get these formulas we have used linear Lagrange interpolations.
Now, from Eq. (\ref{marching_alg}) we obtain the following extrapolation formula:
\begin{widetext}
\begin{equation}
\tilde\gamma^{n}_{j,i}=\frac{1}{1-\delta_i/2\Delta r}\left\{\tilde\gamma_{j,P}+\tilde\gamma_{j,S}-\tilde\gamma_{j,R} + \frac{1}{4}\Delta v (r_Q-r_P+r_S-r_R)\mathcal{H}_C-\frac{1}{2}\frac{\delta_i}{\Delta r}\tilde\gamma^n_{j,i-1}\right\}. \label{algo}
\end{equation}
\end{widetext}
The pending issue to use this formula, that evolves $\hat\gamma$ to the most advanced point, is the evaluation of $\mathcal{H}_C$. We do not show details here, but we make two comments in this respect: i) $\hat U$ and its derivatives are calculated at $n-1/2$ (that is, at the center of the null parallelogram), $j-1/2$ and $i-1$; 
ii) $\hat\gamma$, $\beta$, $M$, $\psi$ and its derivatives are calculated at $n-1/2$, $j$ and $i-1/2$.

{The discretization of the evolution equation (\ref{psievol}) for the scalar field proceeds in the same way.}
\begin{figure}
\includegraphics[width=2.5in,height=1.5in,angle=0]{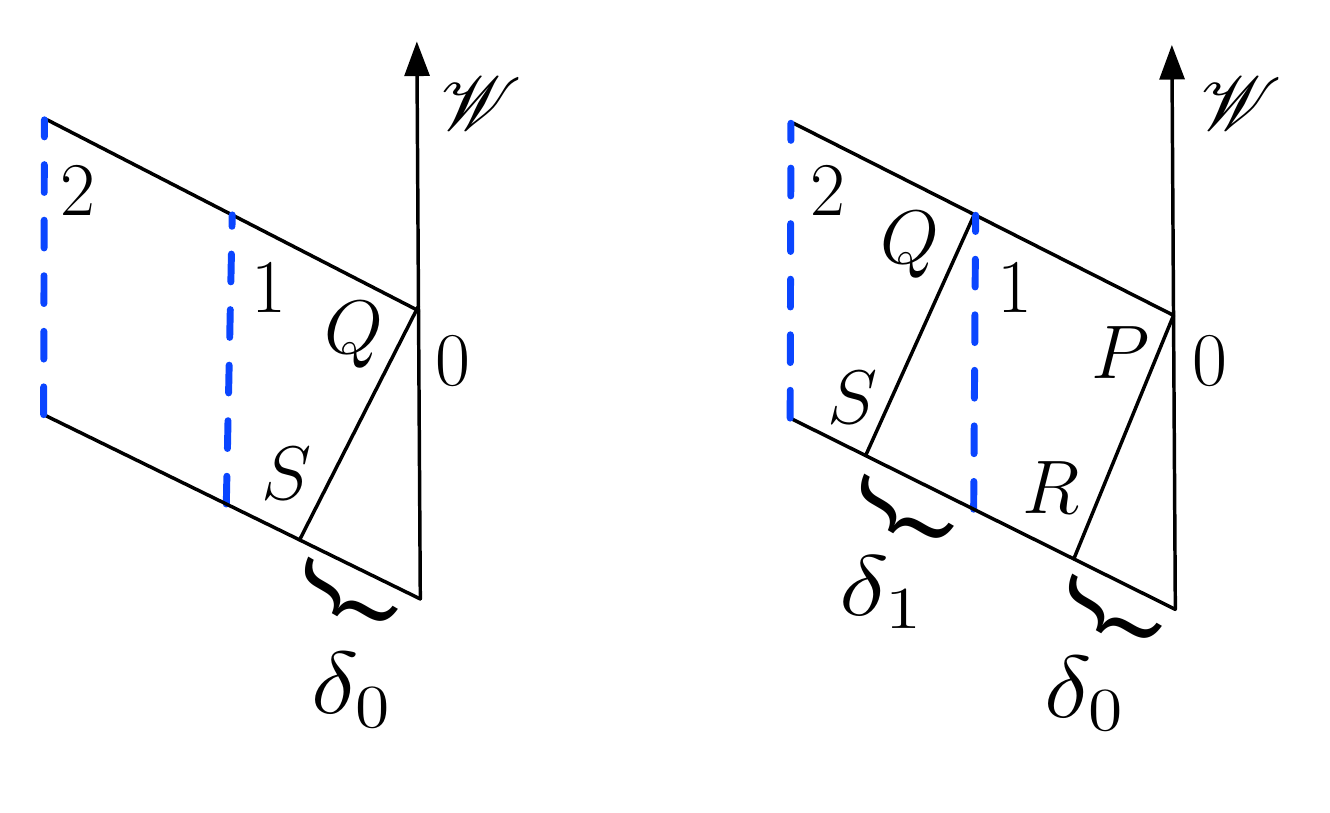}
\caption{Treatment of the first (left) and second (right) points for the ingoing marching algorithm.}
\label{fig:3}
\end{figure}
\begin{figure}
\includegraphics[width=2.5in,height=4.0in,angle=0]{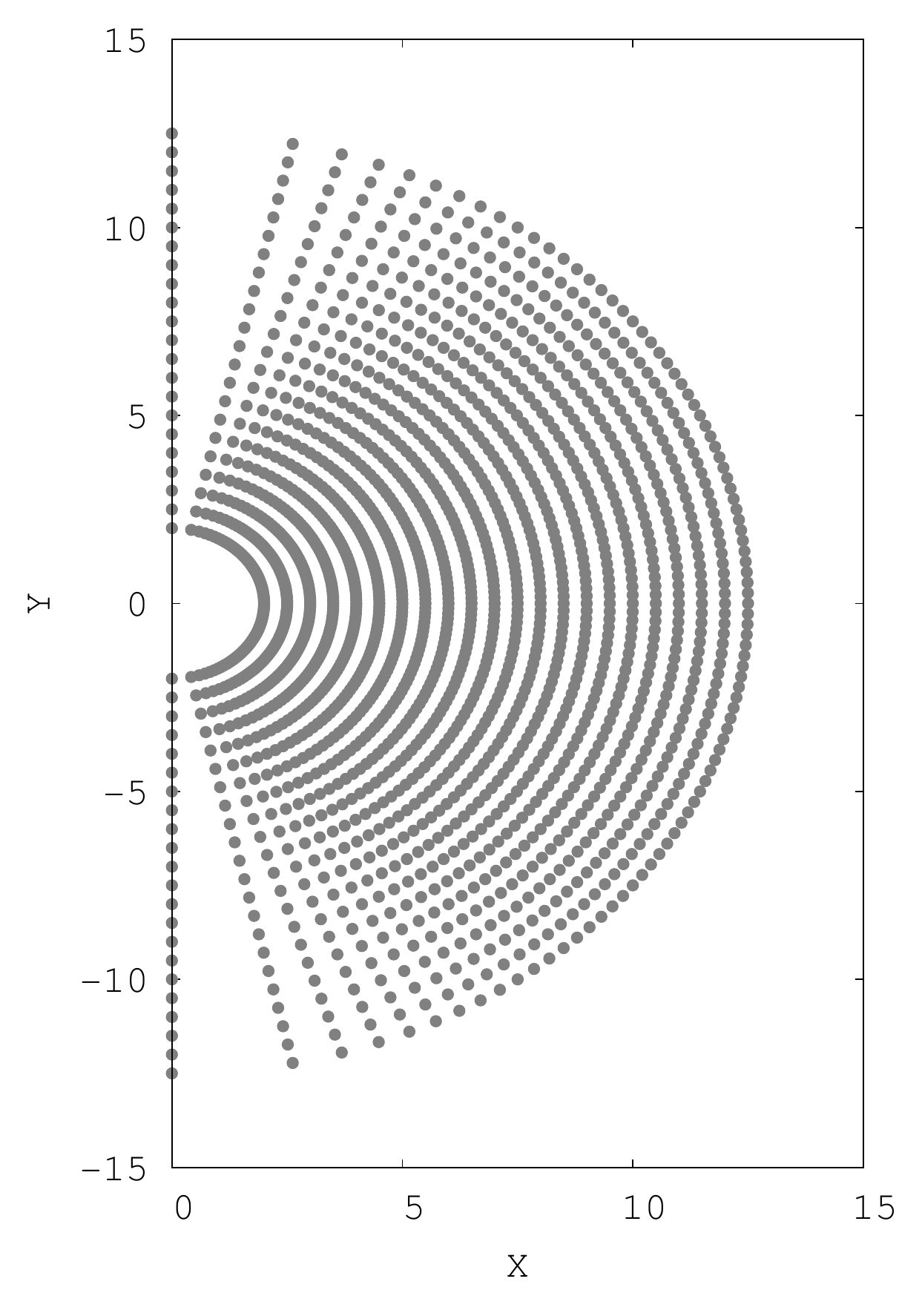}
\caption{Distribution of the grid points on the ingoing light cone using a pseudo Cartesian space $X=r \cos\vartheta,Y=r \sin\vartheta$, where $\vartheta=\pi/2-\theta$. The points $X=0$ denote the axis of symmetry, whereas $Y=0$ denotes the equatorial plane. The grid is uniform in $r$ and $y$ but nonuniform in $\theta$.}
\label{fig:4}
\end{figure}
{\subsubsection{Treatment of the initial-boundary conditions}
The boundary conditions (\ref{bc}) are given at the first and second radial grid points to integrate the hypersurface equations.}
{For these two points we implement the algorithm depicted in Fig. \ref{fig:3} to integrate the evolution equations. The first displacement is
\begin{equation}
\delta_0=\frac{1}{2}\left(1-\frac{2M^{n-1}_{j,0}}{r_0}\right)\Delta v.
\end{equation}
The only point {\it in} the world tube is $S$ at
\begin{equation}
r_S=r_0-\delta_0.
\end{equation}
Thus the interpolation leads us to:
\begin{equation}
\tilde\gamma_{j,S}=\tilde\gamma^{n-1}_{j,0}+\frac{\delta_0}{\Delta r}(\tilde\gamma^{n-1}_{j,1}-\tilde\gamma^{n-1}_{j,0}).
\end{equation}
Then we approximate 
\begin{equation}
\tilde\gamma^n_{j,0}=\tilde\gamma_{j,Q}\leftarrow\tilde\gamma_{j,S}.
\end{equation}
Observe that $Q$ is placed at $i=0$. The second displacement is
\begin{equation}
\delta_1=\frac{1}{2}\left(1-\frac{2M^{n-1}_{j,1}}{r_1}\right)\Delta v.
\end{equation}
to calculate the vertices placed at
\begin{subequations}
\begin{eqnarray}
r_P&=&r_0,\\
r_Q&=&r_1,\\
r_R&=&r_0-\delta_0,\\
r_S&=&r_1-\delta_1.
\end{eqnarray}
\end{subequations}

It can be easily shown that
\begin{subequations}
\begin{eqnarray}
\tilde\gamma_{j,P}&=&\tilde\gamma^n_{j,0},\\
\tilde\gamma_{j,R}&=&\tilde\gamma^{n-1}_{j,0}+\frac{\delta_0}{\Delta r}(\tilde\gamma^{n-1}_{j,1}-\tilde\gamma^{n-1}_{j,0}),\\
\tilde\gamma_{j,S}&=&\tilde\gamma^{n-1}_{j,1}+\frac{\delta_1}{\Delta r}(\tilde\gamma^{n-1}_{j,2}-\tilde\gamma^{n-1}_{j,1}).
\end{eqnarray}
\end{subequations}
Since $Q$ is located at $i=1$, we get
\begin{eqnarray}
\tilde\gamma^n_{j,i}&=&\tilde\gamma_{j,Q}=\tilde\gamma_{j,P}+\tilde\gamma_{j,S}-\tilde\gamma_{j,R}\nonumber\\
&+&\frac{1}{4}\Delta v (r_Q-r_P+r_S-r_R)\mathcal{H}_c.
\end{eqnarray}
For the first two points we approximate $\mathcal{H}_C$ as spherical, which is consistent with the non gravitational radiation condition at the world tube.

{For the evolution equation (\ref{psievol}) we proceed with the first two radial points in the same way, except that instead we use for $\mathcal{H}_C$ the linear approximation for the massless scalar field, in the sense of Sec. II.D. 

{The initial conditions (\ref{icgamma}) and (\ref{icpsi}) are specified on the grid points.}
\subsubsection{Radial and angular grids}
The radial domain goes from $r_\mathscr{B}=1M$ to $r_\mathscr{W}=60M$. Thus, with a radial grid points of $N_r=10^3$, we have $\Delta_r\approx0.06$. For an angular resolution of 90 grid points, $N_y=45$, we cover the angular domain $y\in[-1,1]$ corresponding to $\Delta y\approx 0.02$. The angular grid is uniform in $y$ but not uniform in $\theta$ (see Fig. \ref{fig:4}).  
\subsection{CFL condition}
The stability and convergence of the algorithm depends on the Courant-Friedrichs-Lewy (CFL) condition. Basically the CFL requires that the analytical domain of dependence of the problem be contained in the numerical domain of dependence. This can be satisfied if each grid point at $v_{n-1}$ in Eq. (\ref{algo}) lies on or outside the past characteristic cone of the point $(v_n,r_i,\theta_j,\phi_k)$ to which the fields are being involved. 
We can use the following CFL condition based on a linear analysis of the
evolution system around $r=0$
\begin{equation}dv\le K drdy^2\label{cflinear},\end{equation}
with $K$ of order of one \cite{giw92}. This can be an overly restrictive
condition since the evolution domain does not include $r=0$.}

{Other CFL condition used in \cite{p02} is
\begin{equation}\left[1-\frac{2M}{r}-U^2r^2e^{2(\gamma-\beta)}\right]dv\le 2 dr,
\end{equation}
which supposes $d\theta=d\phi=0$ in the line element.}

{We find the most general treatment of the CFL supposing that the
points $(v-\Delta v, r + \Delta r, \theta, \phi)$ and $(v-\Delta v,r-\Delta r,\theta\pm \Delta\theta,\phi)$ evolve to $(v,r,\theta,\phi)$ on the
null cones open to the past. This leads us to 
\begin{equation}\Delta v \le -2\Delta r\frac{g_{vr}}{g_{vv}}\end{equation}
and

\begin{eqnarray}&&\Delta v\le g_{vv}^{-1}[-(g_{vr}\Delta r \pm g_{v\theta}\Delta\theta)\nonumber \\
&+&
\sqrt{(g_{vr}\Delta r \pm g_{v\theta}\Delta\theta)^2-g_{vv}g_{\theta\theta}\Delta\theta^2}].
\end{eqnarray}
If the fields are strong enough, this constraint take into account the bending of the light--cones.}

In this work we use the CFL condition given by  Eq. (\ref{cflinear}), with $K=0.75$. We observe in practice that $N_y$ has to be at least three times $N_ r$ to guarantee stability.

\subsection{Timing}
Using a grid $N_r\times N_y=256\times90$, an evolution up to $v=1M$ requires 7 minutes (without output) of a 2.4 GHz Intel Core i5.
\begin{table}
\caption{Convergence in amplitude of the 2D EKG code}
\begin{ruledtabular}
\begin{tabular}{ccccc}
$v$&\,\,${\cal Q}_c$ ($10^{-7}$)&\,\,${\cal Q}_m$($10^{-7}$)&
\,\,${\cal Q}_f$($10^{-7}$)&n\\
\hline
0.05&2.108&2.396&2.466&2.04\\
0.10&2.070&2.359&2.430&2.02\\
0.15&2.033&2.323&2.395&2.00\\
0.20&1.997&2.287&2.361&1.99
\end{tabular}
\end{ruledtabular}
\end{table}

\begin{table}
\caption{Convergence in phase of the 2D EKG code}
\begin{ruledtabular}
\begin{tabular}{cccc}
$v$&\,\,${\cal Q}_{cm}$ ($10^{-11}$)&\,\,${\cal Q}_{mf}$($10^{-11}$)&n\\
\hline
0.05&3.207&0.517&2.63\\
0.10&4.457&0.790&2.50\\
0.15&6.489&1.207&2.43\\
0.20&9.286&1.763&2.40
\end{tabular}
\end{ruledtabular}
\end{table}

\subsection{Second order convergence}
We measure the convergence in terms of the norm
\begin{equation}
\mathcal{Q}(v)=\int_{r_\text{in}}^{r_\mathscr{W}} \int_{-1}^1 \psi^2 r^2drdy,
\end{equation}
using the second order accurate Simpson's formula
\begin{equation}
\mathcal{Q}=\sum_{i,j}{\frac{1}{2}[(r\psi)^2_{j,i}+(r\psi)^2_{j,i+1})]}
\Delta r\Delta y.
\end{equation}
For the convergence test we take $r_\text{in}=2.23M$ and $r_\mathscr{W}=60M$, $\lambda=10^{-3}M$ and $\ell=2$. The following grids were used:
\begin{enumerate}[(i)]
\item Coarse, $N_r=125$, $N_y=43$,
\item Medium, $N_r=250$, $N_y=86$,
\item Fine, $N_r=500$, $N_y=172$,
\end{enumerate}
for which $\Delta r$ and $\Delta y$ scale as 4:2:1. Assuming that the quantity $\mathcal{Q}$ behaves as $\mathcal{Q}=a+b\Delta^n$, it can be shown that the convergence rate is
\begin{equation}
n=\log_2\bigg\{\frac{\mathcal{Q}_c-\mathcal{Q}_m}{\mathcal{Q}_m-\mathcal{Q}_f}\bigg\},
\end{equation}
where $\mathcal{Q}_c$, $\mathcal{Q}_m$, $\mathcal{Q}_f$ refer the to computed values of $\mathcal{Q}$ using the coarse, medium and fine grids, respectively \cite{bghlw03}, \cite{bdglrw05}. The results in Table I show that the two-dimensional EKG code is second order convergent in amplitude.

We also measure the convergence in phase. It can be easily shown that the order of convergence in phase is expressed by
\begin{equation}
n=\log_2\bigg\{\frac{\mathcal{Q}_{cm}}{\mathcal{Q}_{mf}}\bigg\},
\end{equation}
where 
\begin{equation}
\mathcal{Q}_{cm}=\int_{r_\text{in}}^{r_\mathscr{W}} \int_{-1}^1 (\psi_c-\psi_m)^2r^2drdy
\end{equation}
and
\begin{equation}
\mathcal{Q}_{mf}=\int_{r_\text{in}}^{r_\mathscr{W}} \int_{-1}^1 (\psi_m-\psi_f)^2r^2drdy
\end{equation}
are calculated at the same grid points and at the same time by subsampling from the fine to the medium grid, and from the medium to the coarse grid. The results in Table II confirm that our code is also second order convergent in phase.

{The same boundary conditions, initial data and marching algorithm were used to calibrate the radial code. The convergence rate in amplitude for the radial code is $1.71$ for the radial grid sizes of $N_r= 126$, $251$, $501$; $1.93$ for $N_r= 501$, $1001$, $2001$, and $2.03$ for $N_r=751$, $1501$, $3001$; all measured at $v=1M$. The convergence rate in phase for the radial code is $2.55$ for the radial grid sizes of $N_r= 126$, $251$, $501$, at $v=1M$. Such a behavior, in both convergences, is shown by the nonlinear 2D EKG code.}
\begin{figure}
\includegraphics[width=2.6in,height=3.1in,angle=0]{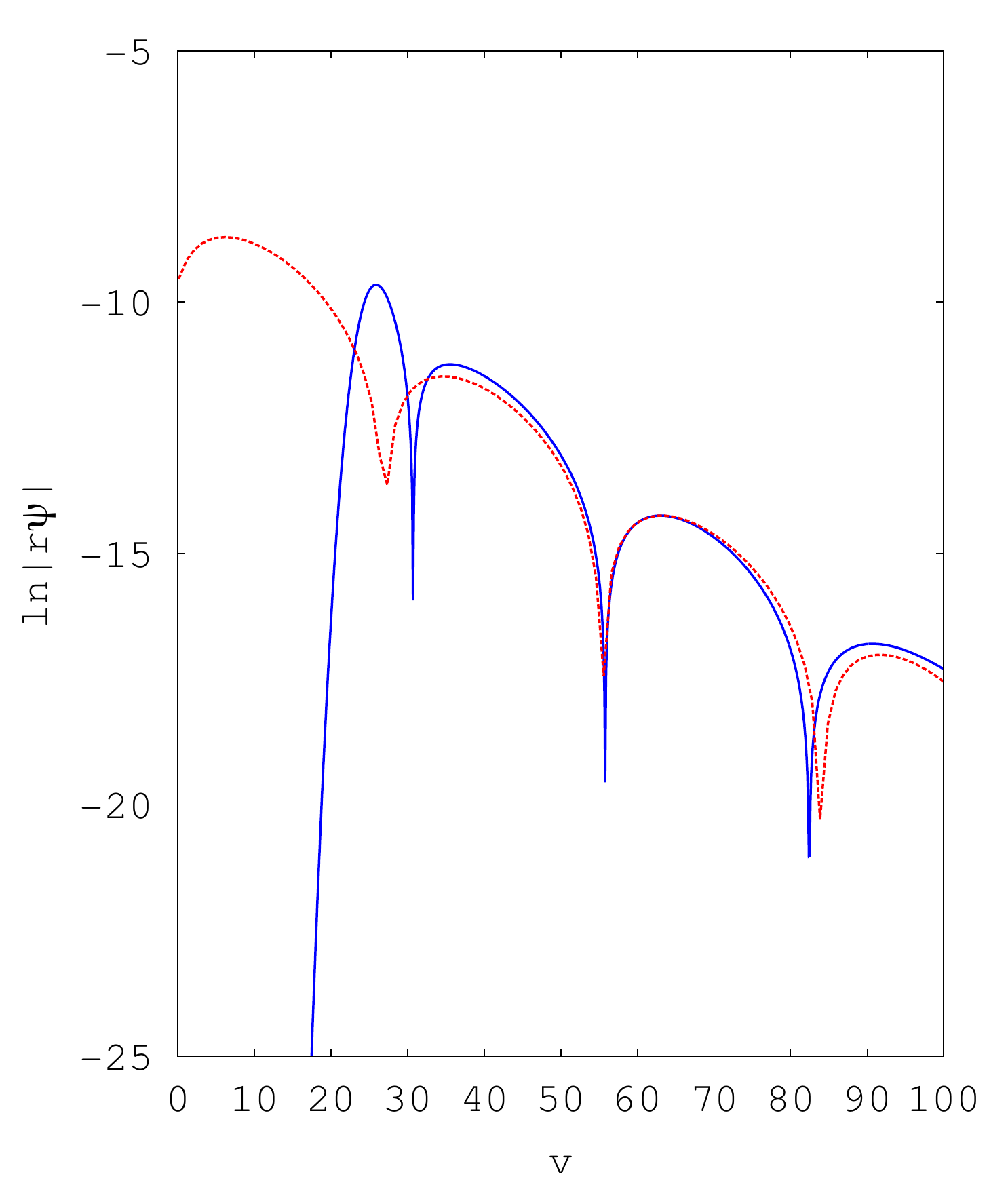}
\caption{Log of the absolute value of the function $r\psi$ at $r=10M$ as a function of Bondi time, showing the quasinormal mode regime oscillations for $\ell=0$.  The solid line is the output for $N_y=90$ and $N_r=2000$; the dashed line is the quasinormal mode extracted from the data.}
\label{fig:qnml0}
\end{figure} 
\begin{figure}
\includegraphics[width=2.6in,height=3.1in,angle=0]{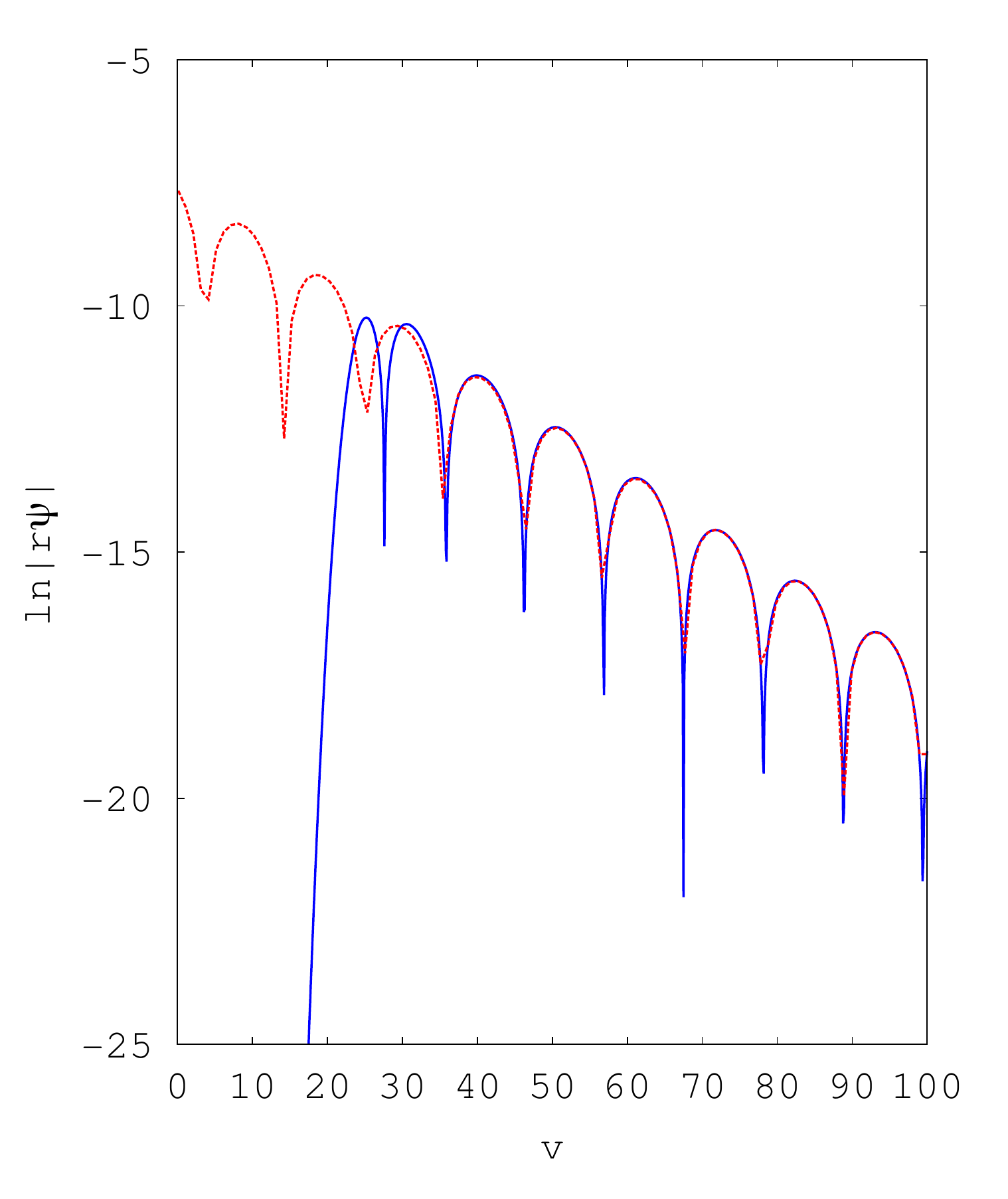}
\caption{Log of the absolute value of the function $r\psi$ at $r=10M$ as a function of Bondi time, showing the quasinormal mode regime oscillations for $\ell=1$.  The solid line is the output for $N_y=90$ and $N_r=2000$; the dashed line is the quasinormal mode extracted from the data.}
\label{fig:qnml1}
\end{figure} 
\begin{figure}
\includegraphics[width=2.6in,height=3.1in,angle=0]{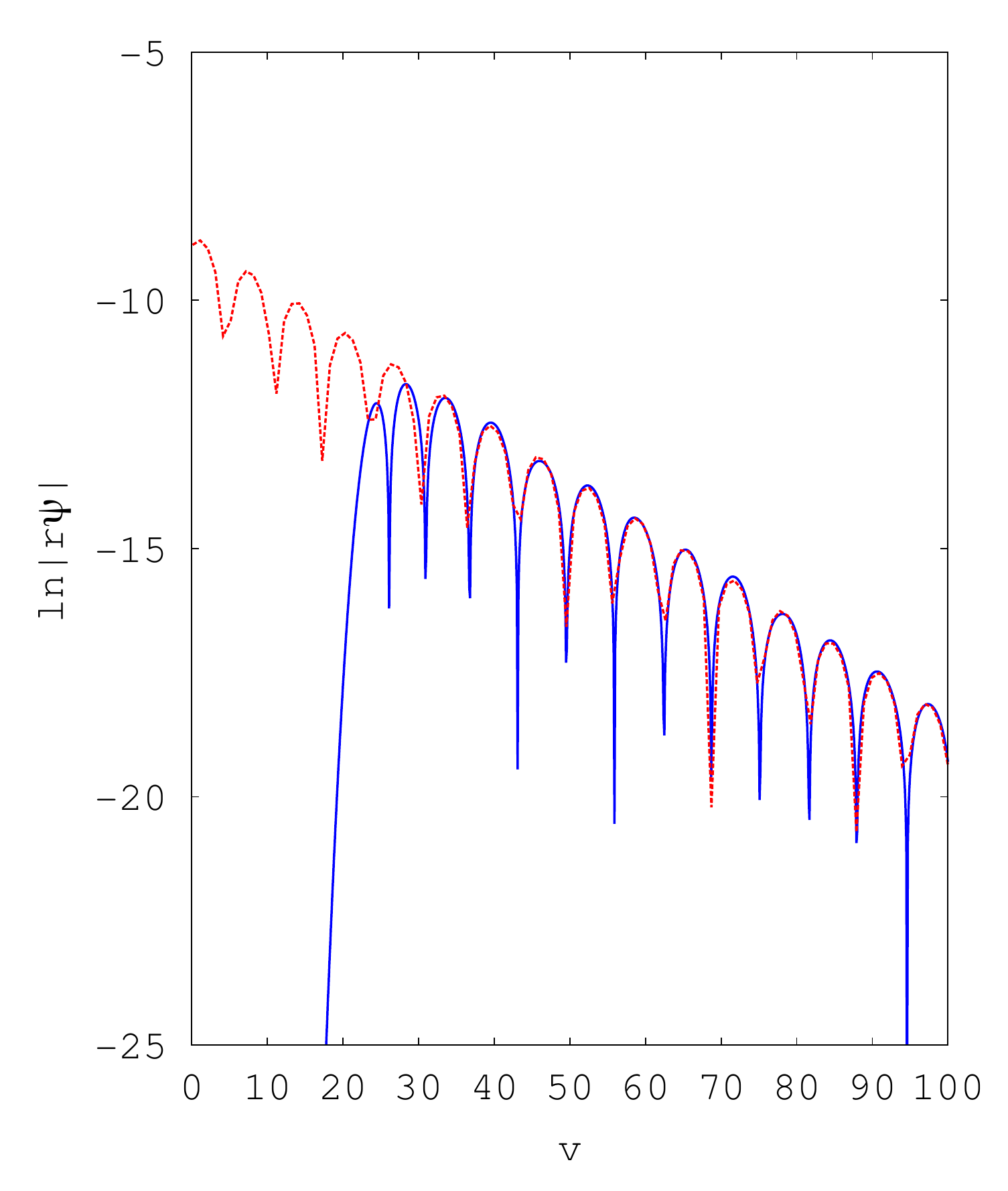}
\caption{Log of the absolute value of the function $r\psi$ at $r=10M$ as a function of Bondi time, showing the quasinormal mode regime oscillations for $\ell=2$.  The solid line is the output for $N_y=90$ and $N_r=2000$; the dashed line is the quasinormal mode extracted from the data.}
\label{fig:qnml2}
\end{figure} 
\begin{figure}
\includegraphics[width=2.6in,height=3.1in,angle=0]{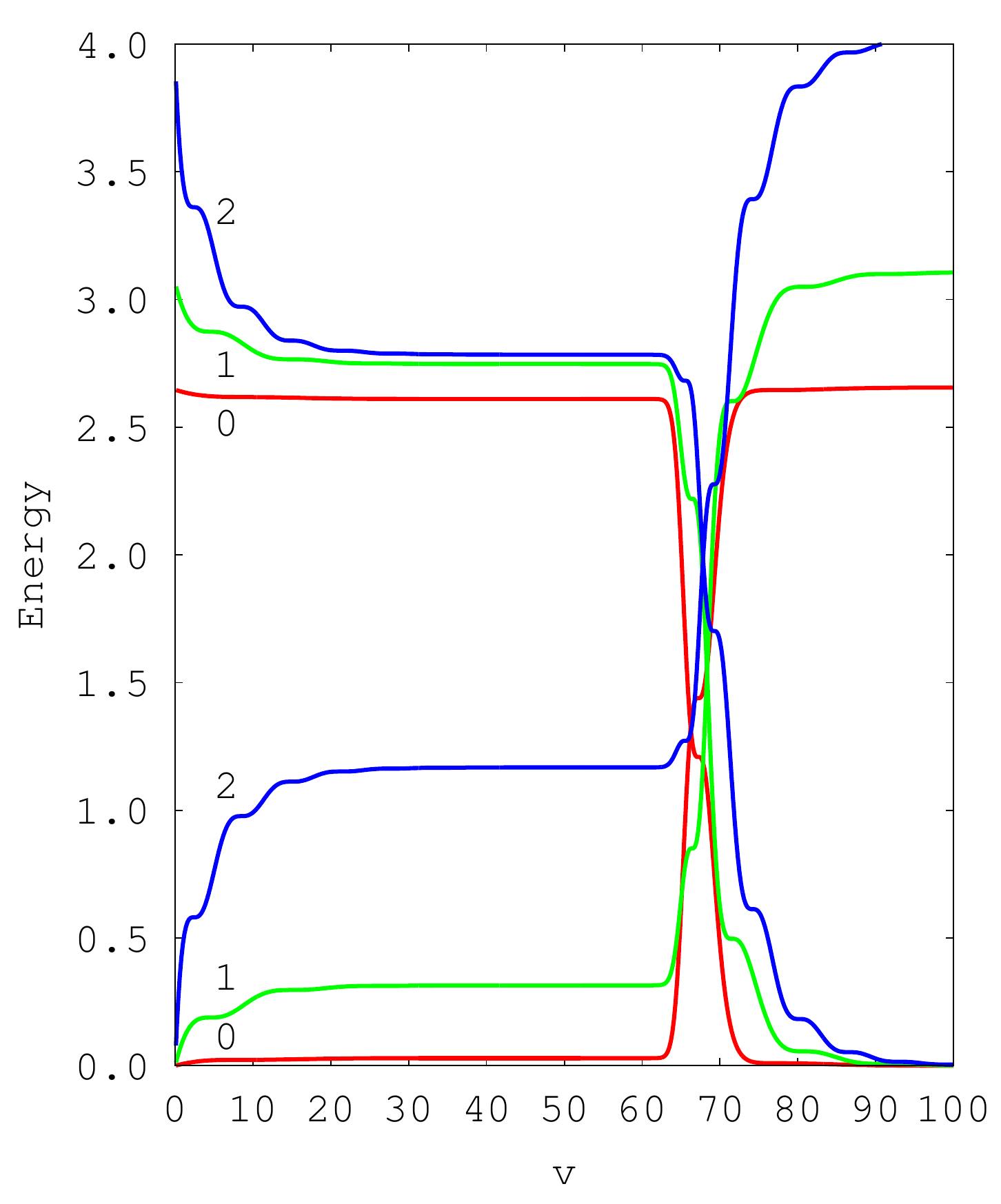}
\caption{Energy conservation (multiplied by $10^8$) as a function of the Bondi time for $\ell=0$ (curve 0); $\ell=1$ (curve 1) $\ell=2$ (curve 2). This calculation was done using the same grid parameters as for Fig. \ref{fig:qnml0}. For each specific $\ell$ the descending curve corresponds to energy given by Eq. (\ref{ten}). The ascending curve corresponds to the algebraic sum of $E_{in}=\int P_{in} dv$ and $E_{out}=\int P_{out} dv$.}
\label{fig:energy}
\end{figure}
\begin{figure}
\includegraphics[width=2.6in,height=3.1in,angle=0]{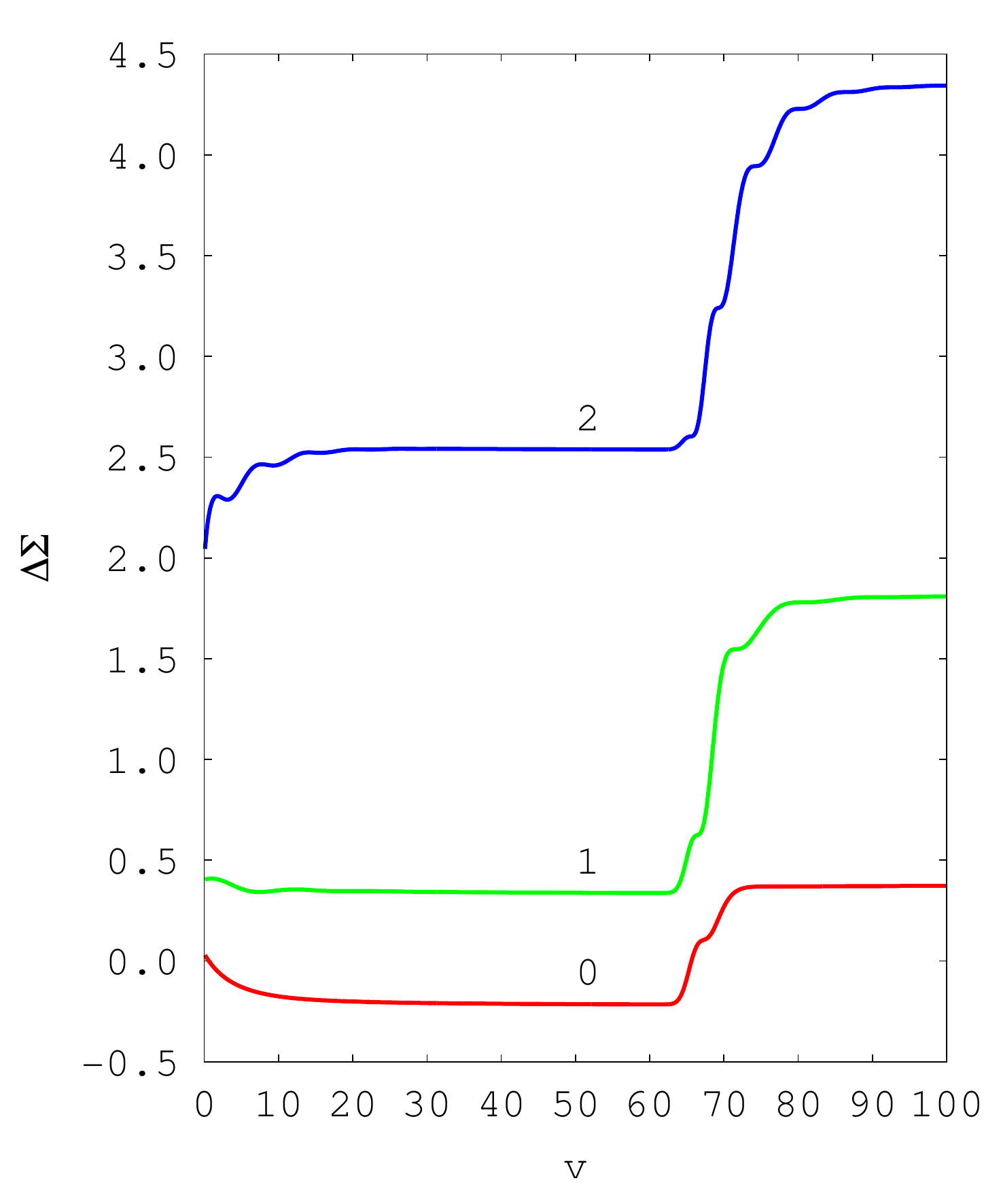}
\caption{Percentage variation in $\Sigma(v)$ with respect to $\Sigma(0)$ as a function of Bondi time for $\ell=0$ (curve 0), $\ell=1$ (curve 1), and $\ell=2$ (curve 2). The graph shows that energy is conserved to within less than $4.5\%$ of the energy content of the initial surface.}
\label{fig:variation}
\end{figure} 
\begin{figure}
\includegraphics[width=2.6in,height=3.1in,angle=0]{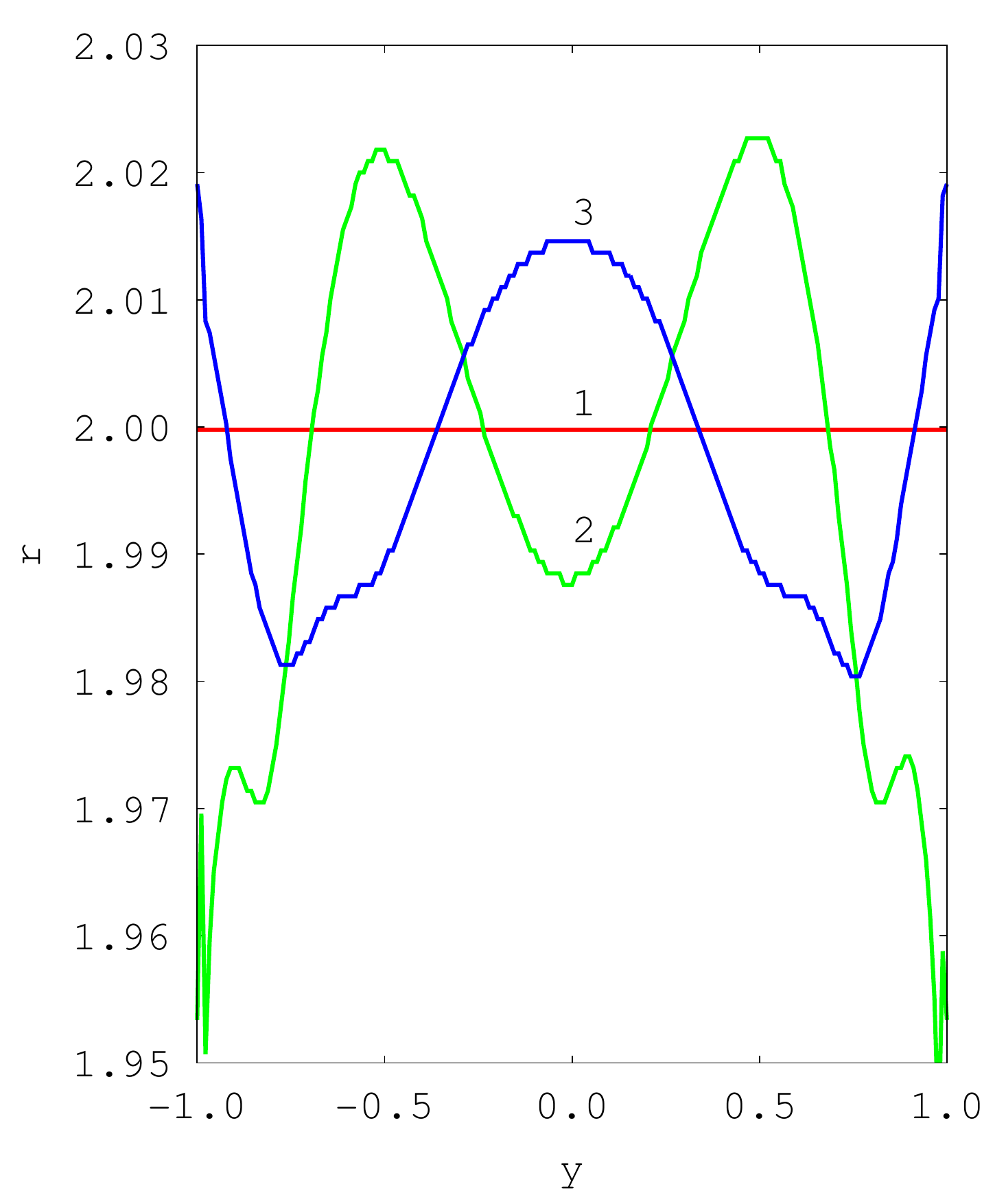}
\caption{Radius of the q-boundary as a function of $y=-\cos(\theta)$ for $\ell=2$, $N_y=90$, $N_r=2000$, $\lambda=0.1$, and different Bondi times $v$: 0 (curve 1); 0.4 (curve 2); 0.8 (curve 3). The MTS is {estimated} using the q-boundary method (see Ref. \cite{gmw97}) with one iteration.}
\label{fig:hrl2}
\end{figure} 
\begin{figure}
\includegraphics[width=1.6in,height=2.1in,angle=0]{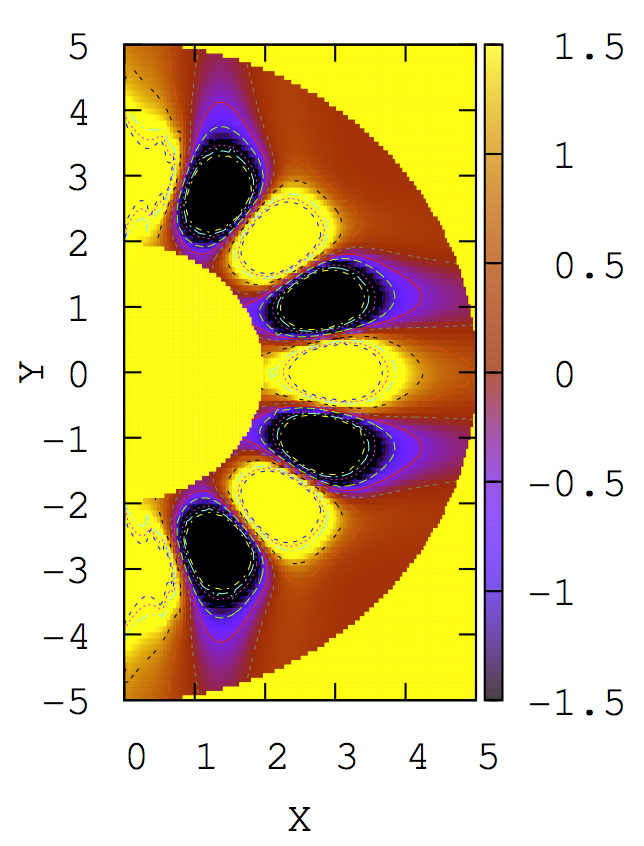}
\includegraphics[width=1.6in,height=2.1in,angle=0]{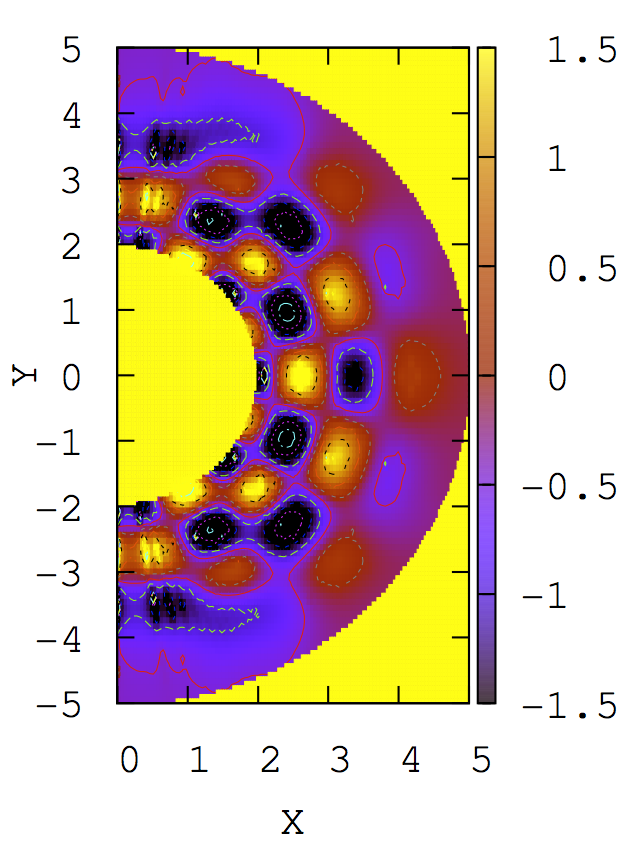}
\includegraphics[width=1.6in,height=2.1in,angle=0]{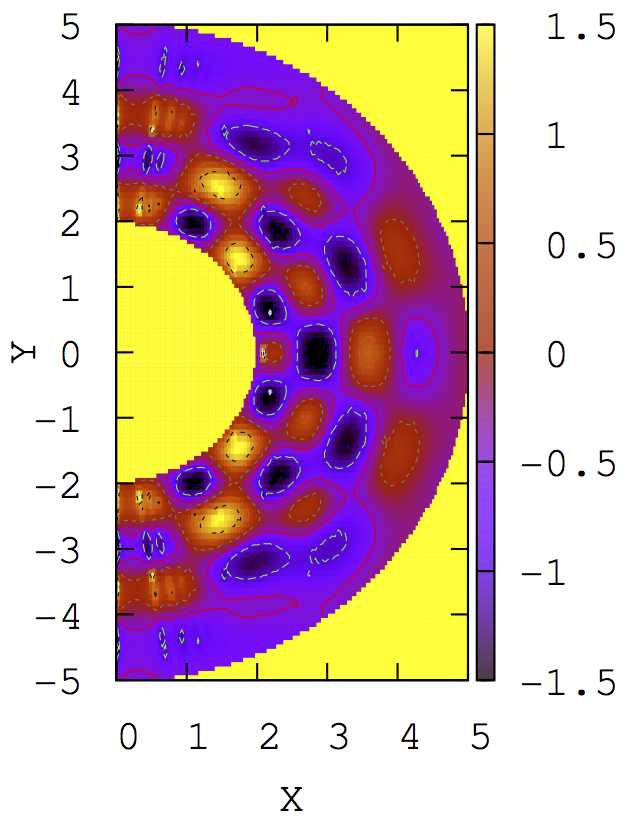}
\includegraphics[width=1.6in,height=2.1in,angle=0]{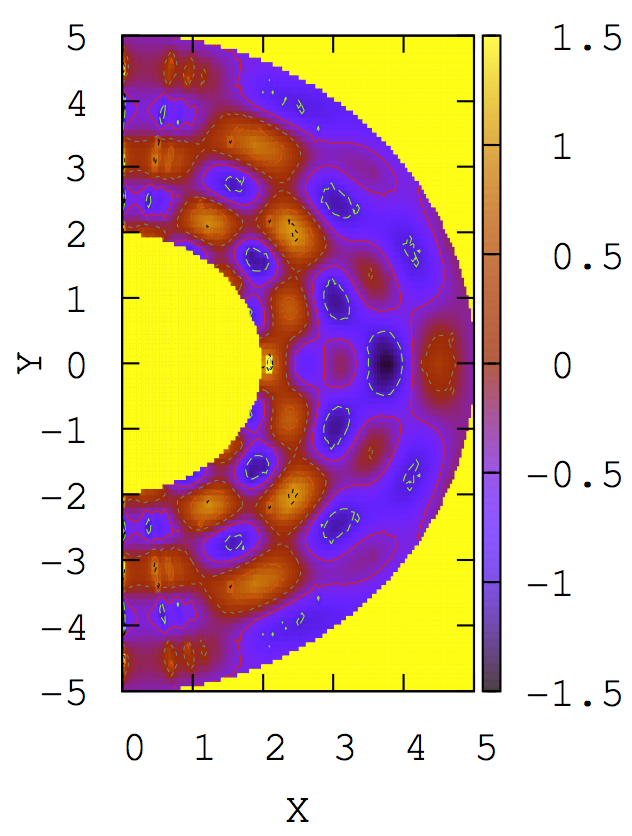}
\caption{{Sequence of snapshots illustrating the evolution of finite amplitude black hole oscillations via an ingoing light--cone approach. The upper left panel shows the scalar field $r\psi$ (multiplied by $10^6$) at $v=0$, representing a localized finite amplitude perturbation of a black hole. The plot coordinates are pseudo--Cartesian $X=r \cos \vartheta$, $Y=r \sin\vartheta$, where $\vartheta=\pi/2-\theta$, hence placing the axis of symmetry along $X=0$ and the equator along $Y=0$. 
Shown are contour levels of the distortion of the light cone. In this case the initial datum is an $l=8$ harmonic.  The next snapshot (upper right panel) shows the evolution of $r\psi$ at $v=2.5M$. The outward propagation of the data is visible but also the change of phase near the horizon. The sequence proceeds with the lower left panel, at time $v=5M$. The evolution continues in the lower right panel ($v=7.5M$). The panel show only the innermost region of the computational domain (about $1/12$ of the total radial extent). Despite the oscillatory nature of these first cycles, examination of the signal shows that only for later oscillations does the black hole spacetime approach the typical quasinormal mode ringing.}}
\label{fig:snapshots}
\end{figure}
\section{Numerical results}
\subsection{QNM}
QNM for the scalar radiation can be read off at finite regions inside the world tube and one particular direction, for instance at $r_{\text{out}}=10M$, $\theta=\pi/4$. For the simulations in this section we use a grid with sizes $N_r=2000$ and $N_y=90$; the initial data correspond to Eqs. (\ref{icgamma}) and (\ref{icpsi}) with $\lambda=10^{-4}$. As in Ref. \cite{gbf07}, to extract the QNM we have used the free software package {\sc harminv} \cite{harminv}, which employs a low storage filter diagonalization method (FDM) for finding the quasinormal modes in a given frequency interval. This software package is based on the FDM algorithm described in \cite{mt97,mt98}. {\sc harminv} provides better accuracy than the fast Fourier transform and is more robust than the least square fit. We do not have other reason for its use, more than its simplicity.

In performing a fit with {\sc harminv} to the scalar field waveforms, sometimes is necessary to factor out, at least approximately, the exponential decay of the signal. This happens when the magnitude of the imaginary part of the frequency (the decay rate) is comparable to the real (oscillatory) part, where the FDM method fails to find a fitting frequency. In those cases, we premultiply the signal by an exponentially increasing function $f = \exp\{\omega_f v\}$, perform the fit with {\sc harminv}, and adjust the frequency obtained accordingly. When an analytic value for the frequency is available, we take its imaginary part as the value for $\omega_f$. In general, when the imaginary part of the frequency is not known, it suffices to use a rough estimate of the decay rate, which can obtained graphically. We also need to decide what range of values of $v$ to use to extract this information. We do this by plotting the signal $r\psi$ and noting when the waveform is clearly periodic with an exponentially decaying envelope. From Figs. \ref{fig:qnml0}, \ref{fig:qnml1} and \ref{fig:qnml2} one can see that the regime starts at $v\approx40M$. We take the end of the fitting interval when the signal no longer appears to be a damped sinusoidal waveform. For initial data of the form as Eq. (\ref{icpsi}), with $\ell=0, 1, 2$, we use {\sc harminv} to extract the frequency, using $v \in [40,100]$ as the fitting interval. For $\ell=0$ the measured frequency is $\omega=0.1102(0.3\%)-0.0971i(3.7\%)$; $\ell=1$ is $\omega=0.2951(0.7\%)-0.0974 i(0.3\%)$; $\ell=2$ is $\omega=0.4896(1.3\%)-0.0970 i(0.2\%)$. Here the values in parentheses indicate the percentage deviation from the value calculated in \cite{k04} via the WKB method to sixth order. We confirm that these results are better than those of Ref. \cite{gbf07}, because we are using the ingoing radial formulation. In fact we read off the frequencies with an error between $0.2\%$ and $3.7\%$. 

\subsection{Energy conservation}
With the same initial data used to get the QNMs, Fig. 8 shows the energy
conservation in the linear regime. It is immediately clear from the graph that the energy contained on the initial slice increases with the value of $\ell$. In all cases energy is conserved within some error depending on resolution. Thus, we can use energy conservation, as well as the results from running the same initial data on the radial code, to debug and calibrate the nonlinear code, as well as to estimate the evolution time needed and its computational requirements. Figure 9 shows the variation in the energy balance $\Delta\Sigma(v)$, defined as the percentage variation in $\Sigma(v)$ relative to the initial value, $\Sigma(v_0)$, i.e.
\begin{equation}
\Delta\Sigma(v)=(\Sigma(v)/\Sigma(0)-1)\times 100
\end{equation}
It can be seen from Fig. 9 that during the simulation, the relative change $\Delta\Sigma(v)$ stays below $0.5\%$ for $\ell=0$, $2.0\%$ for $\ell=1$, and $4.5\%$ for $\ell=2$.
The oscillations observed in the profiles increase with the value of $\ell$ as would be expected. 
%%%%%%%%%%%%%%%%%%%%%%%%%%%%%
{\subsection{MTS and q-boundary}}
We solve Eq. (\ref{Q}) using the bisection method with one iteration. Thus, Fig. 10 shows a rough radius of the q-boundary as a function of $y=-\cos(\theta)$ for $\ell=2$, $N_y=90$, $N_r=2000$, $\lambda=0.1$, at different Bondi times $v$. {Because the q-boundary is always a sphere, Fig. 10 is actually showing
curves with $q=0$, where  the q-boundary seems to be the largest $r$ 
on each curve.} Clearly, in this (early) nonlinear evolution, the MTS develops an angular structure like $\ell=4$. This could be connected with the nonlinear harmonic generation reported in Ref. \cite{p02}, and deserves a future detailed study.

\subsection{Additional test}
%%%%%%%%%%%%%%%%%%%%%%%%%%%%%
In order to get a first glimpse of the type of simulations that our framework enables us to perform, the final calibration check of the code, we select initial data given by Eqs. (\ref{icgamma}) and (\ref{icpsi}), with $\lambda=10^{-4}$, $r_0=3M$, $\sigma=\frac{1}{2} M$, $\ell=8$, and evolve this configuration until $v=7.5M$. The angular grid has the size of $N_y=90$, while the radial grid has $N_r = 2000$ points. This simulation requires 17 hours for each snapshot. Figure \ref{fig:snapshots} displays $r\psi$ as function of $r$ and $\theta$, at $v=0$, $2.5M$, $5M$ and $7.5M$. We assign no particular significance to the selected initial data, other than the fact that its angular complexity provides an excellent test of the code. Despite the oscillatory nature of these first cycles, examination of the signal shows that only for later oscillations does the black hole spacetime approach the typical quasinormal mode ringing.

\section{Conclusions and remarks}
We have extended a computational framework, in the context of the characteristic approach in numerical relativity, to make scalar perturbations of nonrotating black holes. The implementation has 
been used to solve the model problem of a massless scalar field minimally coupled to gravity (the two-dimensional Einstein--Klein--Gordon problem). The procedure is based on the ingoing light cone formulation for an axially and reflection symmetric spacetime. We have shown that our nonlinear code is globally second order convergent in amplitude and phase, and how accurately we can follow the quasinormal mode ringing, for the massless scalar field, including its energy conservation in the linear approximation. As a nonlinear result we show an early MTS evolution developing a higher harmonic. As an additional calibration test we evolve an $\ell=8$ initial harmonic that the code solves quite well, requiring reasonable grid sizes and computing times. 

Currently we are exploring accurately nonlinear effects in the neighborhood of a central black hole. Of particular interest is the study of gravitational waveforms, the marginally trapped surface inside the distorted horizon, and the global energy conservation. {Besides the nonlinear effects (see Ref. \cite{p02}), we are studying the flux of energy across and away the horizon, for the gravitational (and scalar) radiation, including the global energy conservation issues. In this respect, in Ref. \cite{dr11}, global energy conservation was obtained, within some minimized numerical  error, using the Galerkin spectral method to solve the Bondi axial symmetric vacuum problem. Recently, in the spherical symmetric context of the EKG system \cite{b14}, we obtained the global energy conservation in nonlinear and extreme characteristic evolutions. That was possible using the propagation of the descriptor of the asymptotic symmetry and the linkages notion. We are considering such an approach to finite regions using the Galerkin spectral method and the ingoing characteristic formulation.}

{Other future direction includes the application of the present extended framework to a massive and complex scalar field. We envisage the simulation of an initial toroidal boson star which distorts with evolution a spherical black hole.}

\acknowledgments
Thanks to Roberto G\'omez for the facilitation of the two-dimensional vacuum Pittsburgh code, developed for the outgoing characteristic formulation; to Henrique de Oliveira and Carlos Peralta for reading and comments. Also thanks to the Referee who patiently did suggestions to improve the manuscript and to Luis Rosales for hospitality, discussions and encouragement at {\sc unexpo}, Puerto Ordaz, Venezuela. This work was partially financed by Programa de Intercambio Cient\'\i fico, {\sc ula}, M\'erida, Venezuela.

\end{document}